\documentclass[prd,twocolumn,superscriptaddress,floatfix,amsmath,amssymb,amsfonts,nofootinbib,longbibliography]{revtex4-2}

\usepackage{float} 
\usepackage{scalerel}
\usepackage[normalem]{ulem}
\usepackage[english]{babel}
\usepackage{graphicx}
\usepackage{dcolumn}
\usepackage{bm}
\usepackage{blindtext}
\usepackage{verbatim}
\usepackage{relsize}
\usepackage{mathrsfs}
\usepackage{musicography}
\usepackage{amsmath}
\usepackage{blindtext}
\usepackage{cancel}
\usepackage{physics}
\usepackage{epstopdf}
\usepackage{mathtools}
\usepackage{blindtext}
\usepackage{tensor}
\usepackage{color}
\usepackage[usenames,dvipsnames]{pstricks}
\usepackage{epsfig}
\usepackage{pst-grad}
\usepackage{pst-plot} 
\usepackage{hyperref}
\usepackage{verbatim}
\usepackage{slashed}
\usepackage{dsfont}
\usepackage[english]{babel}
\usepackage{amsmath,amssymb,amsthm} 
\usepackage{lipsum}
\usepackage{mathtools} 
\usepackage{tikz}

\newcommand{\m}{\mu}

\newcommand{\w}{\omega}

\newcommand{\mf}{\mathsf}

\newcommand{\ii}{\mathrm{i}}

\renewcommand{\r}{\hat{\rho}}
\newcommand{\s}{\hat{\sigma}}
\newcommand{\tc}[1]{\textsc{#1}}

\newcommand{\tmm}[1]{\textcolor{magenta}{#1}
}

\newcommand{\evv}[1]{\langle #1 \rangle}

\allowdisplaybreaks[1] 

\renewcommand{\r}{\hat{\rho}}

\newcommand{\N}{{{}_{\!\bm N\!}}}

\newcommand{\chapter}[1]{{\center\huge \textbf{#1}}}

\newcommand{\A}{\hat{A}}

\newcommand{\pd}[2]{\frac{\partial {#1}}{\partial {#2}}}

\newcommand{\M}{\mathcal{M}}

\newcommand{\mc}[1]{\mathcal{#1}}
\renewcommand{\AA}{\mathcal{A}}

\newcommand{\hphi}{\hat{\phi}}
\renewcommand{\sp}{ \hat{\sigma}^+}
\newcommand{\sm}{ \hat{\sigma}^-}
\newcommand{\hmu}{\hat{\mu}}

\usepackage{subfiles}

\begin{document}

\title{
Non-perturbative measurements of two-point functions in quantum field theory
}








\author{Sebastian Holm\'en}
\email{sebastianholmen@outlook.com}

\affiliation{Nordita and Stockholm University,
Hannes Alfvéns väg 12, 23, SE-106 91 Stockholm, Sweden}

\author{T. Rick Perche}
\email{rick.perche@su.se}

\affiliation{Nordita,
KTH Royal Institute of Technology and Stockholm University,
Hannes Alfvéns väg 12, 23, SE-106 91 Stockholm, Sweden}

\begin{abstract}
   We present a non-perturbative method through which local probes can access the two-point function of a quantum field within a region of spacetime. By considering a lattice of gapless particle detectors, we identified the probe observables that encode the field's two-point function. We quantify the discrepancies introduced by physical finite-sized interaction regions by performing a spacetime multipole expansion of the smeared two-point function. Our protocol expresses the two-point function entirely in terms of measurable detectors correlations, providing an operational notion of states in QFT.
\end{abstract}


\maketitle

\section{Introduction}

Quantum field theory (QFT) is at the core of our modern description of quantum matter and radiation. In curved spacetimes, however, many constructions that are natural in flat space cease to be unique. In particular, the notion of a preferred vacuum state is generally lost: distinct observers, foliations or coordinates may single out inequivalent mode decompositions and therefore inequivalent vacua. This lack of a distinguished vacuum is a central feature of QFT in curved spacetimes, underlying phenomena such as the Casimir effect~\cite{Casimir1948Attraction,Milton2001CasimirEffect,Bordag2001NewDevelopments}, the Unruh effect~\cite{Unruh1976,Unruh-Wald,takagi,matsasUnruh,UnruhPhilosophers}, and Hawking radiation~\cite{Hawking1975,JormaHawking,ericksonBH}.

At the same time, an operational perspective on QFT does not require a particular choice of vacuum state. Indeed, only three ingredients are required to predict the outcome of any experiment~\cite{daviesAndLewis,Kraus1983StatesEffectsOperations,Busch1996,FewsterVerch}: an initial state for the field, any operations applied to the state, and the observables, whose expectation values are to be computed. In this sense, QFT is fundamentally a framework for assigning probabilities and expectation values to measurements performed in finite spacetime regions, independently of whether the state under consideration can be described as excitations of ``a vacuum''.

From this perspective, the immediate practical question is how one can identify the initial field state. In principle, determining a quantum state would require access to the expectation values of all observables. In QFT this is an exceptionally demanding task, given the infinite-dimensional nature of the algebras of observables of the theory~\cite{Haag,Kadison1963Remarks,Araki1964VonNeumann}. Nevertheless, one expects that by accessing expected values of sufficiently many observables, one would obtain an approximate description of the state. Moreover, the two-point function is enough to fully determine a wide class of states in QFT~\cite{Khavkine2015}. Thus, the question of determining the field's state reduces to the question of measuring the field's two-point function. Equivalently, the state can be estimated if one has access to expected values of sufficiently many pairs of local observables. 

To measure the two-point function, one requires probing the field in different localized regions of spacetime. The most commonly employed tools for this type of local measurements are Unruh-DeWitt (UDW) detectors~\cite{Unruh1976,Unruh-Wald,DeWitt}. These probes have proven to be effective at implementing quantum information protocols in QFT~\cite{Valentini1991,Reznik2003,Pozas-Kerstjens:2015,Jonsson2,collectCalling,Landulfo,ericksonCapacity,KojiCapacity,Ahmadzadegan2021,ericksonCapacity,carol,ericksonNew,boris,quantClass,FullyRelativisticEH}, as well as measuring relevant local properties of quantum fields~\cite{pipo,DanIreneML,geometry,ahmed,perche2024closed,effectOfCurvature2025}. However, usual applications of the model are restricted to the perturbative regime. Indeed, although previous protocols for recovering the two-point function have been put forward~\cite{pipo,geometry}, these have been restricted to the regime of perturbation theory.

In this manuscript we provide a concrete protocol through which one can non-perturbatively obtain the two-point function of an arbitrary state of a real scalar field in a globally hyperbolic background spacetime. The protocol considers a lattice of gapless UDW detectors interacting sequentially with the field, producing a spacetime lattice of interaction regions. By analyzing the correlations between detectors after their interactions, we identify the observables that encode the field's two-point function between different coupling regions. We further quantify the discrepancies between physical (finite-sized) coupling regions and idealized (point-wise) interactions through a multipole expansion. Altogether, the protocol presented here provides both a conceptual and practical framework for extracting field correlations from local measurements.

This manuscript is organized as follows. In Section~\ref{sec:probes in QFT} we set conventions in QFT and review particle detectors as local probes of quantum fields. In Section~\ref{sec:detectors} we compute the final state of $N$ gapless detectors interacting with a quantum field, as well as relevant detector observables non-perturbatively. Section~\ref{sec:extract} contains the main results of our manuscript, outlining a protocol that can be used to recover a field's two-point function using a lattice of gapless detectors. In Section~\ref{sec:error estimation} we estimate the discrepancies between the point-wise two-point function and its smeared version based on a multipole expansion on the size of the coupling regions. Section~\ref{sec:examples} discusses examples of the multipole expansion and the protocol for recovering a field's two-point function with different states in Minkowski spacetime. The conclusions of our work are presented in Section~\ref{sec:conclusions}.

\section{Local Probes in QFT}\label{sec:probes in QFT}

In this section we will set the conventions used throughout the manuscript. Subsection~\ref{sub:QFT} discusses local quantum field theory and Subsection~\ref{sub:UDW} contains a brief review of particle detector models, which will be essential to formulate our explicit protocol to extract the two-point function of quantum fields.

\subsection{Quantum field theory and the Wightman function}\label{sub:QFT}


Since we will be concerned with local measurements in QFT, we will use the language of algebraic quantum field theory~\cite{Haag,kasiaFewsterIntro,Khavkine2015} to describe a real scalar field in a 3+1 globally hyperbolic spacetime $\mathcal{M}$. In this language, a quantum field is described by an association $\mc{A}$ of causally convex subsets $\mc{O} \subseteq \M$ to $\ast$-algebras of observables, \mbox{$\mc{A} : \mc{O} \mapsto \mc{A}(\mc{O})$}. In the specific case of a Klein-Gordon field with equation of motion $P\phi = 0$  \footnote{Here $P$ is a linear hyperbolic differential operator}, the local algebras $\mathcal{A}(\mathcal{O})$ are generated by an identity element and symbols $\hat{\phi}(f)$, where $f\in C^\infty_0 (\mathcal{O})$. The following properties of the elements $\hat{\phi}(f)$ define the algebraic structure of the association $\mathcal{A}$:
\begin{enumerate}
    \item
    $\hphi(\alpha f + g) = \alpha \hphi(f) + \hphi(g)
    ~~ \forall f,h \in C^\infty_0(\M), 
    ~~ \alpha\in\mathds{C}$,
    \item 
    $(\hphi(f))^\dagger = \hphi(f^*),~~ \forall f \in C^\infty_0(\M)$,
    \item 
    $\hphi(f+Ph) = \hphi(f) ~~ \forall f,h \in C^\infty_0(\M)$,
    \item 
    $[\hphi(f),\hphi(g)] = \ii E(f,g)
    ~~ \forall f,h \in C^\infty_0(\M)$,
\end{enumerate}
where $E(f,g)$ is the smeared causal propagator, a bi-distribution with kernel defined as the difference $E(\mf x, \mf x')\equiv G_R(\mf x, \mf x')-G_R(\mf x',\mf x)$, where $G_R(\mf x,\mf x')$ is the retarded propagator. $G_R(\mf x,\mf x')$ tells us how a Dirac delta source for the field at $\mf x'$ propagates forward in spacetime. 


The algebra generators $\hat{\phi}(f)$ can be thought of as smeared field observables. Indeed from property 1 above we see that $\hat{\phi}(f)$ is linear in $f$, so that it can be thought as
\begin{equation}\label{eq:phiDist}
    \hphi(f) = \int \dd V \hphi(\mf x)f(\mf x),
\end{equation}
such that 
$\hphi: C^\infty_0(\M) \mapsto\AA(\M)$
is an ``operator-valued distribution" with kernel $\hat{\phi}(\mf x)$. Although this kernel is not a well-defined operator, $\hat{\phi}(\mf x)$ is what is usually called the quantum field. Equation~\eqref{eq:phiDist} also motivates the remaining properties 2-4. For instance, property 2 ensures that the field operator is self-adjoint $\hat{\phi}^\dagger(\mf x) = \hat{\phi}(\mf x)$, property 3 imposes the equations of motion at the level of the kernel $P\hat{\phi}(\mf x) = 0$, and property 4 imposes the covariant canonical commutation relations.

In this formulation states are defined as linear functionals that map operators to their corresponding expected values, $\w: \AA(\M) \mapsto \mathds{C}$ satisfying the normalization condition $\w(\mathds{1}) = 1$ and positivity $\w(\A^\dagger\A) \geq 0 ~\forall\A\in\AA(\M)$. Given that the association $f\mapsto\w(\hphi(f))$ is linear, it can be thought of as a distribution $W_1:f\mapsto \w(\hphi(f))$ such that
\begin{equation}
    \w(\hphi(f)) =
    \int \dd V W_1(\mf x)f(\mf x) \equiv W_1(f).
\end{equation}
Similarly, the association $(f,g) \mapsto \w(\hphi(f)\hphi(g))$ defines a bi-distribution
\begin{equation}
    W_2(f,g) \equiv 
    \w(\hphi(f)\hphi(g)) =
    \int \dd V \dd V' W_2(\mf x,\mf x')f(\mf x)g(\mf x').
\end{equation}
In this context the kernel $W_1(\mf x)$ defines the one-point function of the state $\omega$, and $W_2(\mf x, \mf x')$ is its two-point function. It is common to refer to the two-point function as the Wightman function, denoting its kernel by $W(\mf x, \mf x')$ and the application to functions by $W(f,g)$. The Wightman function is particularly descriptive when the state $\omega$ is quasifree (a zero-mean Gaussian state), where the odd-point functions vanish, and all even-point functions can be computed from $W(\mf x, \mf x')$. Importantly, a wide class of states in QFT are quasifree, including vacuum and thermal states. These examples are then fully characterized by their two-point functions.



We end this subsection by stating a relevant decomposition of the Wightman function in terms of its symmetric and anti-symmetric parts. Using the identity \mbox{$\hat{\phi}(f)\hat{\phi}(g) = \tfrac{1}{2}\{ \hphi(f),\hphi(g) \}+\tfrac{1}{2}[\hphi(f),\hphi(g)]$}, the anti-commutator $[\hphi(f),\hphi(g)] = \ii E(f,g)$, and defining the Hadamard function $H_\w(f,g) \equiv \w\big(\{ \hphi(f),\hphi(g) \}\big)$, the Wightman function can be written as a sum of symmetric ($H_\omega)$ and anti-symmetric ($E$) bi-distributions:
\begin{equation}\label{eq:W=H-iE}
    W(f,g) =
    \frac{1}{2}H_\w(f,g) + \frac{\ii}{2}E(f,g).
\end{equation}
Note that the second term is independent of the state $\w$, since the causal propagator depends only on the classical equation of motion. The Hadamard term then encodes all the state dependence of the two-point function and completely defines quasifree states.

\subsection{Unruh-DeWitt detectors as local probes}\label{sub:UDW}


The only way to access local information from a quantum field is to use local probes that interact with it in finite regions of spacetime.
The paradigmatic choice for these local probes is the UDW detector model~\cite{Unruh1976,DeWitt}, able to balance simplicity with the ability to capture the essential characteristics of matter-field interactions~\cite{eduardo,Nicho1,richard,neutrinos,antiparticles,pitelli}. This model has also proven to be a fundamental tool for describing numerous relativistic quantum information protocols, such as entanglement harvesting~\cite{Valentini1991,Reznik2003,Pozas-Kerstjens:2015,Ahmadzadegan2021,carol,ericksonNew,boris,quantClass,FullyRelativisticEH}, quantum energy teleportation~\cite{teleportation,HottaEntanglement,teleportation2014}, quantum collect calling~\cite{Jonsson2,collectCalling}, among others~\cite{Farming,KojiCapacity,phil}. More relevant for this manuscript, UDW detectors can also be used to recover the two point function of quantum fields in the regime of weak interactions~\cite{pipo,geometry}. In this subsection we will review the two-level UDW model and set the conventions that will be used throughout the manuscript.



A two-level UDW detector coupled to a scalar field consists of a qubit undergoing a given trajectory in spacetime. To define the model, let $\mf z(\tau)$ be a timelike trajectory parametrized by its proper time $\tau$, and $\tau(\mf x)$ a local extension of the proper time around the curve. The free Hamiltonian generating time evolution with respect to $\tau$ is prescribed as
\begin{equation}
    \hat{H}_{\tc{d}} = \Omega~\sp\sm  = \Omega \ket{e}\!\!\bra{e},
\end{equation}
with $\Omega$ being the energy gap of the detector, $\{\ket{g},\ket{e}\}$ being its ground and excited states and $\{\sm,\sp\}$ being the standard SU(2) raising and lowering operators. The coupling of the qubit to the field is defined by the interaction Hamiltonian density
\begin{equation}\label{UDW hamiltonian}
    \hat{h}_I(\mf x) =
    \lambda \Lambda(\mf x)
    (e^{\ii \Omega\tau(\mf x)}\sp+e^{-\ii \Omega\tau(\mf x)}\sm)\hat{\phi}(\mf x),
\end{equation}
written in the interaction picture, where $\lambda$ is a coupling constant and $\Lambda(\mf x)$ is the spacetime smearing function localized around $\mf{z}(\tau)$, defining the spacetime region where the detector couples to the field. If one can find coordinates $(\tau, \xi^i)$, such that $\xi^i$ parametrize the rest surfaces where $\tau$ is constant (such as Fermi normal coordinates~\cite{poisson}), it is common to assume that $\Lambda(\mf x) = \chi(\tau)f(\bm \xi)$. This corresponds to assuming that the shape of the detector (in space) remains unchanged throughout its interaction with the field~\cite{Schlicht,us,us2,generalPD}. We call $\chi(\tau)$ the switching function and $f(\bm \xi )$ the spatial smearing function. In these coordinates 
\begin{equation}
    \begin{aligned}
        \hat{h}_I(\tau,\bm \xi) =
        \lambda\chi(\tau)f(\bm \xi)
        (e^{\ii\Omega \tau } \hat{\sigma}^+
        +
        e^{-\ii\Omega \tau } \hat{\sigma}^-)\hphi(\mf x).
    \end{aligned}
\end{equation}

Despite the merits of UDW detectors, their application is often restricted to the perturbative regime $\lambda\ll 1$. One exception is when the energy levels of the detector are degenerate, corresponding to $\Omega = 0$\,\footnote{Note that even in this case, it still makes sense to consider ``ground'' and ``excited'' states, as one could reintroduce an energy gap between two standard states for preparation and measurement, but not while the detector interacts with the field.}. In this case the interaction Hamiltonian for the detector becomes simply
\begin{equation}\label{eq:gapless h}
    \hat{h}_I(\mf x) =
    \lambda
    \Lambda(\mf x)    
    \hmu
    \hphi(\mf x),
\end{equation}
where we defined the detector monopole, $\hmu\equiv\sp+\sm$. In the gapless model, one can use the Magnus expansion~\cite{magnus} to obtain the final state of the detector non-perturbatively~\cite{nogo}. This result has been since generalized to the case where 2 detectors interact with the field~\cite{Landulfo,perche2024closed} and to $N$ detectors in~\cite{phil}. In the next section we will use these results to formulate a protocol to extract the Wightman function between two finite spacetime regions.

\section{\textit{N} gapless UDW detectors}\label{sec:detectors}

In this section we will compute the evolved initial state of an $N$ gapless detector system interacting with the field in arbitrary spacetime regions. We will also obtain explicit expressions for expected values of relevant operators, which will we will then relate to the field's two-point function in Section~\ref{sec:extract}. 


Consider $N$ gapless UDW detectors interacting with the field in spacetime regions $\Lambda_i(\mf x)$ with interaction Hamiltonians of the form of Eq.~\eqref{eq:gapless h}. The total interaction Hamiltonian density is then
\begin{equation}
    \hat{h}_I(\mf x) = 
    \lambda \sum_{i=1}^N \Lambda_i(\mf x)  \hat{\mu}_i\hphi(\mf x),
\end{equation}
where we denoted each detector's monopole moment by $\hat{\mu}_i = \hat{\sigma}_i^+ + \hat{\sigma}_i^-$ and we assumed that all detectors share the same coupling constant $\lambda$.

 The dynamics of the detectors-field systems is given by the time evolution operator 
\begin{equation}\label{eq U definition}
    \hat{U} = \mc T\!\exp
    \left(
    -\ii \int\dd V \hat{h}_I(\mf x)
    \right)
\end{equation}
with $\mc T\!\exp$ being the time ordered exponential\footnote{Note that for gapless detectors, the time evolution operator is uniquely defined and does not depend on a specific choice of time parameter, unlike the case of gapped detectors~\cite{us,us2}.}. For $N$ gapless detectors we can compute $\hat{U}$ non-perturbatively using the Magnus expansion~\cite{Landulfo,perche2024closed,phil}. In Appendix.~\ref{app:evolution} we compute the final state of the detectors in terms of smeared field propagators when the detectors all start in the ground state and the field starts in an arbitrary state $\omega$.
After tracing out the field degrees of freedom we find the final detector state to be
    \begin{equation}\label{eq:detector state}
        \begin{aligned}
            \r_\tc{d} 
            &= \frac{1}{2^N}\sum_{\bm \mu,\bm \mu'}
            e^{
            \frac{\ii }{2}
            \sum_{i<j}
            (\mu_i'\mu_j'-\mu_i\mu_j)\Delta_{ij}
            }
            e^{
            \frac{\ii}{2}\sum_{i} \sum_{j}\mu_i'\mu_j 
            E_{ij}
            } \\[-3mm]
            &~~~~~~~~~~~~~~~~
            \times e^{
            -\frac{1}{2}
           \sum_{i}
            \sum_{j}
            (\mu_i-\mu_i')(\mu_j-\mu_j')W_{ij}
            }
            ~
            \ket{\bm \mu}\!\!
            \bra{\bm \mu'},
        \end{aligned}
    \end{equation}
where $\bm \mu = (\mu_1,...,\mu_N)$, the sum is over \mbox{$\mu_i = \pm$} for all $i$, we denote the eigenvectors of $\hat{\mu}_i$ by $\ket{\pm_i}$, and we introduce the notation 
\begin{equation}  
\begin{aligned}H_{ij} &\equiv \lambda^2 H(\Lambda_i,\Lambda_j), &&& E_{ij} &\equiv \lambda^2 E(\Lambda_i,\Lambda_j),\\
W_{ij} & \equiv\lambda^2W(\Lambda_i,\Lambda_j),&&& \Delta_{ij}&\equiv\lambda^2\Delta(\Lambda_i,\Lambda_j).
\end{aligned}
\end{equation}
Here $\Delta(\mf x, \mf x')$ denotes the symmetric propagator, with $\Delta(\Lambda_i,\Lambda_j)=G_R(\Lambda_i,\Lambda_j )+G_R(\Lambda_j,\Lambda_i)$, and $G_R(\Lambda_i,\Lambda_j)$ denotes the retarded Greens function smeared over the interaction regions $\Lambda_i$ and $\Lambda_j$.

We are interested in the expected values of observables acting on the final detector state. 
As we will see later, the field's two-point function becomes encoded in correlated observables between pairs of detectors. For this reason, we will only consider expected values of pairs of operators on different detectors. For $\alpha \in \{0,x,y,z\}$, we denote the corresponding sigma matrix acting on the $i$th qubit by  $\s_\alpha^{(i)}$, so that we compute expected values of the form $\langle{\s^{(i)}_\alpha\s^{(j)}_\beta}\rangle$. In Appendix~\ref{app:ev} we show that
\begin{widetext}
    \begin{equation}\label{eq:general ev useful}
        \begin{aligned}
            \langle \s^{(i)}_\alpha\s^{(j)}_\beta
            \rangle_{\r_\tc{d}}
            &=
            \sum_{\bm \mu,\bm \mu'}
            \frac{1}{2^N}
            \bra{\mu_i'\mu_j'}
            \s^{(i)}_\alpha\s^{(j)}_\beta
            \ket{\mu_i\mu_j}
            ~
            e^{
            -\frac{1}{2}\left((\mu_i-\mu_i')^2 W_{ii} + (\mu_i-\mu_i')(\mu_j-\mu_j')H_{ij}+(\mu_j-\mu_j')^2 W_{jj} \right)
            }
            \\
            &
           \quad\quad\quad\quad\quad\quad\quad\quad\quad
            \times
            e^{
            \ii\left(
            (\mu_i'-\mu_i)
            \tilde{G}_i(\bm \mu)
            +
            (\mu_j'-\mu_j)
            \tilde{G}_j(\bm \mu)
            +
            \tfrac{1}{2}(\mu_i'\mu_j'-\mu_i\mu_j)\Delta_{ij}
            +
            \tfrac{1}{2}(\mu_i'\mu_j - \mu_j'\mu_i)
            E_{ij}
            \right)},
        \end{aligned}
    \end{equation}
\end{widetext}
where have introduced the symbols  
\begin{equation}  
\begin{aligned}\Delta_{i,j}&\equiv\lambda^2\Delta(\Lambda_i,\Lambda_j), &&& E_{ij} &\equiv \lambda^2 E(\Lambda_i,\Lambda_j),\\
W_{ij} & \equiv\lambda^2W(\Lambda_i,\Lambda_j),&&& H_{ij} &\equiv \lambda^2 H(\Lambda_i,\Lambda_j),
\end{aligned}
\end{equation}
as well as
\begin{equation}
    \tilde{G}_i(\bm \mu)
\equiv
\frac{1}{2}
\sum_{\substack{k=1\\k\neq i,j}}\m_k(E_{i k} + \Delta_{i k}) = 
\sum_{\substack{k=1\\k\neq i,j}}\m_kG_R(\Lambda_i,\Lambda_k),
\end{equation}
which encodes the total propagation of information from all detectors $k$ to detector $i$ conditioned on the state $\ket{\bm \mu}$.
The specific expected values can be obtained by inserting the corresponding value of 
$\bra{\mu_i'\mu_j'}
 \s^{(i)}_\alpha\s^{(j)}_\beta
\ket{\mu_i\mu_j}$ in Eq.~\eqref{eq:general ev useful}. We compiled a list of all nonzero expected values in Appendix~\ref{Appendix List of nonzero expected values}.

\section{A protocol for measuring the Hadamard function}\label{sec:extract}

The fact that quasifree states in QFT are defined by their two-point functions implies that unless one can measure $W(\mf x, \mf x')$ between \textit{every} pair of events, one cannot infer in which state the field is at. In this section we will show how gapless particle detectors can be used to measure the two-point function of a field between pairs of events within a spacetime lattice. More precisely, we will recover the smeared Hadamard function between pairs of spacetime regions, which can then be combined with the causal propagator (see Eq.~\eqref{eq:W=H-iE}) to reconstruct the two-point function. In the limit where the detectors interaction regions are sufficiently localized, we effectively obtain the two-point function between the center of the regions.


The first step to recover the Wightman function within a spacetime lattice will be to define a modified UDW model, that can couple to the field multiple times, while encoding information about each coupling region in independent degrees of freedom. In this model \textit{each detector} consists of $N$ qubits, thus being described in the Hilbert space $(\mathbb{C}^{2})^{\otimes N}$. We assume that the detector undergoes a timelike trajectory $\mf z(\tau)$ that crosses\footnote{This assumption implies that there exists a set of proper times $\{\tau_i\}$ such that $\mf z(\tau_i) = \mf x_i$ for each $i$, and we further assume that $\tau_i<\tau_{i+1}$, implying that $\mf x_{i}$ is always in the causal past of $\mf x_{i+1}$.} a set of events $\mf x_i$. Each degree of freedom described within a $\mathbb{C}^2$ subspace then couples to the field in a region defined by $\Lambda_{\mf x_i}(\mf x)$ centered at $\mf x_i$. Explicitly, each detector's interaction with the field is modelled by a Hamiltonian density of the form
\begin{equation}
    \hat{h}_I'(\mf x) =
    \lambda
    \sum_{i=1}^N
    \Lambda_{\mf x_i}(\mf x)
    \hmu_{i}
    \hphi(\mf x),
\end{equation}
where $\hat{\mu}_i = \hat{\sigma}_i^+ + \hat{\sigma}_i^-$ is the detector's monopole moment at the $i$th interaction, and $\hat{\sigma}^\pm_i$ are the raising and lowering operators on the $i$th $\mathbb{C}^2$ factor of the detector's Hilbert space. Effectively, this sequential detector model consists of $N$ independent UDW models, each of which couples to the field at a later region. 

To recover the Wightman function between events in a spacetime lattice, we will consider a three dimensional lattice of sequential detectors that each interact with the field at different times. For instance, if the lattice contains $N^3$ different detectors, each interacting $N$ times, this results in a four dimensional lattice of $N^4$ different interaction regions defined by functions $\Lambda_{\mf x_i}(\mf x)$, $i \in N^{\times4}$. Thus, $\{\mf x_i\}$ defines a spacetime lattice. We can measure the observables~\eqref{eq:general ev useful} between pairs of qubits, from which we will be able to extract the Hadamard function. Indeed, in the case where $\Lambda_{\mf x_i}$ and $\Lambda_{\mf x_j}$ are spacelike separated, we can write (see Appendix~\ref{Appendix Extracting the Hadamard function in genereal})
\begin{equation}\label{eq:Hab simple}
    \begin{aligned}
        H_{ij}
        =
        \frac{1}{2}\text{arctanh}\left(\frac{
        \langle\s^{(i)}_y\s^{(j)}_y\rangle}{\langle\s^{(i)}_z\s^{(j)}_z\rangle}\right),
    \end{aligned}
\end{equation} 
expressing the smeared two-point function entirely in terms of measurable detector correlations. 

In contrast, when the pair of detectors $(i,j)$ is causally connected, the expected values in~\eqref{eq:Hab simple} pick up additional contributions due to their causal influence through retarded propagators. In Appendix~\ref{Appendix Extracting the Hadamard function in genereal} we show that these causal contributions are encoded in the term 
\begin{equation}
    C_{ij}
    =
    \frac{1}{2}
    \sum_{\substack{k=1 \\ k\neq i,j}}^N
    \text{arctanh}
    \left(
    \frac{\langle\s^{(i)}_y\s^{(k)}_x\rangle}{\langle\s^{(i)}_z\rangle}
    \frac{\langle\s^{(k)}_x\s^{(j)}_y\rangle}{\langle\s^{(j)}_z\rangle}
    \right),
\end{equation}
so that, in general, the Hadamard function can be written as
\begin{equation}\label{eq:HC}
     H_{ij}
        =
        \frac{1}{2}\text{arctanh}\left(
        \frac
        {\langle\s^{(i)}_y\s^{(j)}_y\rangle}{\langle\s^{(i)}_z\s^{(j)}_z\rangle}
        \right) - C_{ij}.
\end{equation}

At this stage we can obtain $H(\Lambda_{\mf x_i},\Lambda_{\mf x_j})$ between two regions. Knowing the field's classical equations of motion, we can also compute $E_{ij}$, yielding the Wightman function between the regions $W_{ij} = \lambda^2 W(\Lambda_{\mf x_i},\Lambda_{\mf x_j})$ through Eq.~\eqref{eq:W=H-iE}. Notice that if one has access to $W_{ij}$ for any arbitrary regions $\Lambda_{\mf x_i}$ and $\Lambda_{\mf x_j}$, this fully determines a quasifree state. 

In practice, a setup for measuring $W_{ij}$ would involve preparing a specific field state, setting up the spacetime lattice of interactions and collecting statistics for the expected values in Eq.~\eqref{eq:HC}. This setup would then have to be repeated (with the same field state) until enough statistics are collected this way, yielding an approximation of $W_{ij}$ for the corresponding interaction regions. In each implementation, we would need to have access to $4$ expected values involving only detectors $i$ and $j$ ($\langle\s^{(i)}_z\s^{(j)}_z\rangle$, $\langle \s^{(i)}_y\s^{(j)}_y\rangle$, $\langle\s^{(i)}_z\rangle$ and $\langle\s^{(j)}_z\rangle$)  and $2(N-2)$ expected values involving $i$ or $j$ and the other detectors ($\langle\s^{(i)}_y\s^{(k)}_x\rangle$ and $\langle\s^{(k)}_x\s^{(j)}_y\rangle$), resulting in  $4+2(N-2)=2N$ total expectation values. For each pair of causally disconnected detectors the data required is much lower, requiring access to only 2 quantities (see Eq.~\eqref{eq:Hab simple}).

There are two important advantages that the protocol presented here has when compared to previous setups that use localized probes to extract two-point functions in QFT~\cite{pipo,geometry}. The first advantage is theoretical. Due to the fact these previous works have considered gapped UDW detectors, only perturbative results were considered. This limitation makes it so that the final states obtained in~\cite{pipo,geometry} are singular in the limit where the interaction regions are spacetime Dirac deltas, $\Lambda_{\mf x_i}(\mf x) \to \delta_{\mf x_i}(\mf x)$. This happens because in this limit terms such as $W_{ii}$ and $H_{ii}$ diverge to infinity. In contrast, in our non-perturbative approach, Eqs.~\eqref{eq:detector state} and~\eqref{eq:general ev useful} are regular in the limit $W_{ii},H_{ii}\to\infty$. However, in this limit the correlations $H_{ij}$ become negligible compared to the noise introduced by the local coupling with the field, yielding a fully dephasing evolution for the qubits. Nevertheless, taking the limit of $\ell\to 0$ in Eqs.~\eqref{eq:Hab simple} and~\eqref{eq:HC} enables one to, at least in theory, recover the Wightman function point-wise. 

The second advantage is due to the fact that non-trivial energy gaps hinder the detector's ability to access the field's two-point function. For detectors with energy gap $\Omega$ one recovers the Wightman function from expectation values that depend on terms of the form $W(\Lambda_{\mf x_i}^-,\Lambda_{\mf x_j}^+)$, where $\Lambda_{\mf x_i}^\pm(\mf x) = e^{\pm\ii \Omega \tau(\mf x)}\Lambda_{\mf x_i}(\mf x)$. In the limit of small energy gaps $\Lambda_{\mf x_i}^\pm(\mf x) \approx (1\pm \ii \Omega \tau(\mf x))\Lambda_{\mf x_i}(\mf x)$, implying that in the limit $\Lambda_{\mf x_i}(\mf x) \to \delta_{\mf x_i}(\mf x)$,
\begin{equation}
    W(\Lambda_{\mf x_i}^-,\Lambda_{\mf x_j}^+) \approx \big(1 - \ii \Omega \tau(\mf x_i) +\ii \Omega \tau(\mf x_f)\big)W(\mf x_i,\mf x_j),
\end{equation}
introducing additional imaginary dependencies on $\Omega$ and the proper times of the detectors. Moreover, for large $\Omega$ compared to the temporal support of the interaction, the added oscillations  $e^{\pm \ii \Omega \tau(\mf x)}$ make it so that \mbox{$W(\Lambda_{\mf x_i}^-,\Lambda_{\mf x_j}^+) \not\approx W(\mf x_i,\mf x_j)$}. The gapless model does not introduce any unnecessary oscillations to the smeared propagators, allowing one to approximate the Wightman function between two spacetime events, even when the interaction regions are not arbitrarily small. In the next section we will quantify the difference between the smeared two-point functions obtained when using physical finite-sized detectors compared to $W(\mf x_i,\mf x_j)$.







\section{Estimating error from using physical detectors}\label{sec:error estimation}

At this stage, we developed a method for recovering the Hadamard function between spacetime regions. Although we showed that our setup is regular in the spacetime delta limit, physical probes are not pointlike and cannot couple to the field instantaneously. In this section we will quantify the difference between the two-point function $W(\Lambda_{\mf x_i},\Lambda_{\mf x_j})$, obtained when using physical probes, and $W(\mf x_i,\mf x_j)$, recovered in the limit $\Lambda_{\mf x_i}(\mf x) \to \delta_{\mf x_i}(\mf x)$.


Our strategy will be to use a multipole expansion to express the smeared Wightman function over two regions $\Lambda_{\mf x_i}$ and $\Lambda_{\mf x_j}$ in terms of $W(\mf x_i,\mf x_j)$. The multipole expansion of the smeared Wightman function around the events $\mf x_i$ and $\mf x_j$ reads
\begin{equation}\label{Wightman expanded}
        W(\Lambda_{\mf x_i},\Lambda_{\mf x_j}\!) \!=\!\!\!
        \sum_{\smash{\substack{{}\\n=0 \\ m=0}}}^\infty\!
        \frac{1}{n!m!}
        W_{\mu_1...\mu_n\nu_1'...\nu_m'}\!(\mf x_i,\mf x_j)
        \Lambda^{\mu_1...\mu_n}_{\mf x_i}
        \Lambda^{\nu_1'...\nu_m'}_{\mf x_j}\!,
\end{equation}
where primed indices denote tensors at $\mf x_j$ and unprimed indices denote tensors at $\mf x_i$, we denote
\begin{align}
    &W_{\mu_1...\mu_n\nu_1'...\nu_m'}
    (\mf x_i, \mf x_j)
    \\
    &~~~~~~~~~~~~
    \equiv\pd{}{x^{\mu_1}}...\pd{}{x^{\mu_n}}
    \pd{}{x^{\nu_1'}}...\pd{}{x^{\nu_m'}}
    W(\mf x, \mf x')\eval_{\substack{\mf x = \mf x_i \\ \mf x'= \mf x_j}},\nonumber
\end{align}
and the multipole moments of $\Lambda_{\mf x_i}(\mf x)$ around $\mf x_i$ are
\begin{equation}\label{Lambda moment}
    \begin{aligned}
        \Lambda^{\mu_1\dots\mu_n}_{\mf x_i}
        \equiv
        \int \dd V \sigma^{\mu_1}(\mf x,\mf x_i)
        \dots\sigma^{\mu_n}(\mf x,\mf x_i)
        \Lambda_{\mf x_i}(\mf x),
    \end{aligned}
\end{equation}
where $\sigma^\mu(\mf x,\mf x_i)$ denotes the generalized separation vector\footnote{$\sigma^\mu(\mf x, \mf x_i)$ corresponds to the initial vector of a geodesic starting at $\mf x_i$ and reaching $\mf x$ such that $\sigma^\mu\sigma_\mu$ is the geodesic squared spacetime separation between $\mf x$ and $\mf x_i$. This notation comes from derivatives of Synge's world function (see e.g.~\cite{poisson}).} between $\mf x_i$ and $\mf x$ and analogous expressions hold for the multipoles of $\Lambda_{\mf x_j}(\mf x)$, $\Lambda_{\mf x_j}^{\nu_1'...\nu_n'}$. Notice that in the expansion~\eqref{Wightman expanded}, the first term ($n=m=0$) corresponds to spacetime Dirac delta interactions, while the remaining terms will be corrections due to the finite size of the detectors. For instance, if the interaction regions have a characteristic size $\ell$ in both space and time, Eq.~\eqref{Wightman expanded} takes the form $W(\Lambda_{\mf x_i},\Lambda_{\mf x_j}) = W(\mf x_i, \mf x_j) + \mathcal{O}(\ell)$.

We can quantify the corrections due to finite size by prescribing an explicit shape for the spacetime smearing functions $\Lambda_{\mf x_i}(\mf x)$. To do so in a general hyperbolic spacetime, we will use Riemann normal coordinates $\mf y^\alpha=(\mf x-\mf x_i)^\alpha$ centered at $\mf x_i$ to prescribe the respective spacetime smearing functions as
\begin{equation}\label{eq:LambdaRiemann}
    \Lambda_{\mf x_i}(\mf x)
    =
    \frac{
    e^{\frac{-(\mf y^0)^2-\mf y^i\mf y^j\delta_{ij}}{2\ell^2}}}
    {(2\pi)^2\ell^4}.
\end{equation}
This defines $\Lambda_{\mf x_i}(\mf x)$ as an effective geodesic 4-ball in spacetime with characteristic size $\ell$ in both the temporal and spatial directions.

With the spacetime smearing functions of~\eqref{eq:LambdaRiemann}, we can explicitly compute the corresponding multipole moments by performing the integration in Eq.~\eqref{Lambda moment} in Riemann normal coordinates. To this end, we notice that $\mf y^\mu = \sigma^\mu(\mf x, \mf x_i)$ and expand the volume element as
\begin{equation}\label{dV}
    \dd V = \dd^4\mf y\sqrt{-g}
    \approx
    \dd^4\mf y \Big(1-\frac{1}{6}R_{\alpha\beta}(\mf x_i)\mf y^\alpha \mf y^\beta
    \Big),
\end{equation}
where $R_{\alpha\beta}(\mf x_i)$ is the Ricci tensor at $\mf x_i$.
In Appendix~\ref{Appendix Details for error estimation} we compute the moments of the smearing functions, obtaining
\begin{align}\label{error estimate}
        W(\Lambda_i,\Lambda_j)&=
        W(\mf x_i,\mf x_j)
        \\
        &
        -\frac{\ell^2}{6}W(\mf x_i, \mf x_j)(\delta^{\alpha\beta}R_{\alpha\beta}(\mf x_i)+\delta^{\alpha'\beta'}R_{\alpha'\beta'}(\mf x_j))\nonumber
        \\
        &+
        \frac{\ell^2}{2}
        (W_{\mu\nu}(\mf x_i, \mf x_j)
        \delta^{\mu\nu}\!+\!W_{\mu'\nu'}(\mf x_i, \mf x_j)
        \delta^{\mu'\nu'})\nonumber
        \\
        &\quad\quad\quad\quad\quad\quad\quad\quad\quad\quad\quad\quad\quad\quad\quad\,\,+\order{\ell^4}.\nonumber
\end{align}
We see that, with this choice of highly symmetric spacetime smearing functions with vanishing dipole moments, the effect of using physical detectors scales as the square of the extension of their interaction region in spacetime $\ell$. 
In the next section we will explore explicit examples of this expansion.

\section{Explicit Examples in Minkowski spacetime}\label{sec:examples}


In this section we will study examples of different field states in Minkowski spacetime in the context of the multipole expansion of Section~\ref{sec:error estimation} as well as the protocol of Section~\ref{sec:extract} in the case of quasifree states. These examples will allow us to explicitly see the effect of finite regions in measuring the field's two-point function, as well as the limits of the multipole expansion discussed in Section~\ref{sec:error estimation}. In the examples we will consider comoving inertial detectors in Minkowski spacetime, with interaction regions centered at events $\mf x_i = (t_i,\bm x_i)$ written in inertial coordinates $(t,\bm x)$ as
\begin{equation}\label{eq:Smearing region}
    \Lambda_{\mf x_i}(\mf x)
    =
    \frac{
    e^{\frac{-(t-t_i)^2-|\bm x - \bm x_i|^2}{2\ell^2}}}
    {(2\pi^2)^2\ell^4},
\end{equation}
where $\ell$ controls the spacetime extension of the couplings. We will use interactions of this form to study the smeared Wightman function of the Minkowski vacuum, thermal states, one-particle wavepackets and coherent states.


\subsection{Minkowski vacuum}

The real part of the vacuum two-point function in Minkowski spacetime can be written, in inertial coordinates, as
\begin{equation}\label{Wightman Minkowski}
    W_0(\mf x, \mf x')=
    \frac{1}{4\pi^2\big(-(t-t')^2+(\bm{x}-\bm{x}')^2\big)}
\end{equation}
whenever $-(t-t')^2+(\bm{x}-\bm{x}')^2 \neq 0$. When smeared against test functions $\Lambda_{\mf x_i}$ and $\Lambda_{\mf x_j}$ of the form~\eqref{eq:Smearing region} the result can be computed in closed form (see e.g.~\cite{perche2024closed}). When $\Delta t_{ij} = |t_i- t_j|= 0$ and $\Delta x_{ij} = |\bm x_i - \bm x_j| = s$, we obtain
\begin{equation}
    W^{(\ell)}(\Lambda_{\mf x_i},\Lambda_{\mf x_j}) = \frac{(s/\ell)^{-1}e^{- \frac{s^2}{8\ell^2}}}{8 \sqrt{2}\pi^{3/2}\ell^2}\text{erfi}\left(\frac{s/\ell}{2\sqrt{2}}\right),
\end{equation}
and when the interactions are only separated in time with $\Delta t_{ij} = |t_i- t_j| = s$ and $\Delta x_{ij} = |\bm x_i - \bm x_j| = 0$, 
\begin{equation}
    W^{(\ell)}(\Lambda_{\mf x_i},\Lambda_{\mf x_j}) = \frac{1}{16\pi^2\ell^2} - \frac{(s/\ell)\, e^{-\frac{s^2}{8 \ell^2}}}{32 \sqrt{2}\pi^{3/2}\ell^2}\text{erfi}\left(\frac{s/\ell}{2\sqrt{2}}\right).
\end{equation}

In this case the asymptotic expansion of Section~\ref{sec:error estimation} reduces to a small $\ell$ expansion, resulting in the leading order expansion (See Appendix~\ref{Appendix Details for error estimation} for details)
\begin{equation}
    W^{(2)}\!(\Lambda_{\mf x_i},\Lambda_{\mf x_j})\! = \!W_0(\mf x_i,\mf x_j)\!\left(\!1 + \ell^2\! \bigg(\!\frac{6\Delta t_{ij}^2 + 2\Delta x_{ij}^2}{(- \Delta t_{ij}^2 + \Delta x_{ij}^2)^2}\!\bigg)\!\!\right)\!.
\end{equation}
We see that, as expected, as $\ell\to 0$, the behaviour of $W^{(2)}(\Lambda_{\mf x_i},\Lambda_{\mf x_j})$ becomes close to that of $W_0(\mf x_i, \mf x_j)$, and the  corrections compared to the pointlike case grow as $\ell$ increases. Since Minkowski spacetime is translation invariant, the relative size of the corrections are determined by the dimensionless ratio $\ell/s$, where 
\mbox{$s^2 \equiv(\mf x_i - \mf x_j)^\mu(\mf x_i - \mf x_j)_\mu$}. 

\begin{figure}[h!]
    \centering
    \includegraphics[width=1\linewidth]{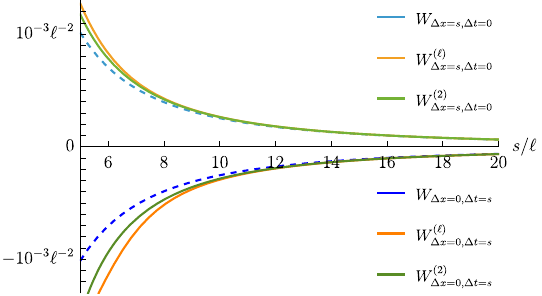}
    \caption{The pointlike (blue), smeared (orange), and leading order multipole expansion (green) of the vacuum two-point function as a function of the ratio of the interactions space and time separation $s$ by their characteristic size in spacetime $\ell$.}
    \label{fig:mincvac}
\end{figure}
In Fig.~\ref{fig:mincvac} we plot the vacuum Wightman function between events, the two-point function smeared against functions of the form~\eqref{eq:Smearing region}, and its leading order multipole expansion, all as a function of $s/\ell$. The positive values correspond to interaction regions with $\Delta t_{ij} = 0$ and the negative values correspond to $\Delta x_{ij} = 0$.  We see that all two point functions acquire a similar behaviour as the ratio $s/\ell$ becomes large, corresponding to the regime where the detectors are able to accurately recover the two-point function and where the multipole expansion is valid. In contrast, as $s/\ell$ decreases, smeared detectors cannot accurately capture the behaviour of the Wightman function.

\subsection{Thermal state}

In this subsection we will consider thermal states in Minkowski spacetime. In inertial coordinates, the two-point function of a thermal state with inverse temperature $\beta$ can be written as
\begin{align}
    W^{(\beta)}(\mf x, \mf x') = \frac{1}{8\pi\beta |\Delta \bm x|}\bigg(&\text{coth}\left(\tfrac{\pi}{\beta}(|\Delta\bm x| + \Delta t)\right)\\&+\text{coth}\left(\tfrac{\pi}{\beta}(|\Delta \bm x| - \Delta t)\right)\bigg)\nonumber.
\end{align}
Performing the multipole expansion of Section~\ref{sec:error estimation}, we find simple expressions for the leading order corrections due to the size of Gaussian probes. Explicitly, when the center of the interactions are such that $\Delta \bm x = 0$, we find
\begin{equation}
    W^{(2)}(\Delta t) =- \frac{1}{4 \beta^2 \sinh^2\big(\tfrac{\pi}{\beta}\Delta t\big)} - \frac{\pi^2 \ell^2(2 + \cosh{}\big(\tfrac{2\pi}{\beta}\Delta t\big))}{\beta^4 \sinh^4\big(\tfrac{\pi}{\beta}\Delta t\big)},
\end{equation}
and when $\Delta t = 0$, we find
\begin{equation}
    W^{(2)}(|\Delta \bm x|) = \frac{\coth\big(\tfrac{\pi}{\beta}|\Delta \bm x|\big)}{4  \beta|\Delta \bm x|} + \frac{\pi \ell^2\coth\big(\tfrac{\pi}{\beta}|\Delta \bm x|\big)}{|\Delta \bm x| \beta^3 \sinh^2\big(\tfrac{\pi}{\beta}|\Delta \bm x|\big)}.
\end{equation}

\begin{figure}[h!]
    \centering
    \includegraphics[width=\linewidth]{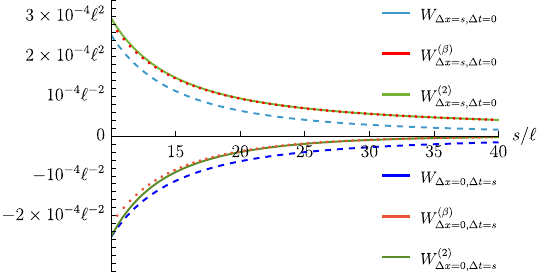}
    \caption{The two-point function for the Minkowski vacuum (blue), a thermal state at inverse temperature $\beta = 50\ell$ (red), and its multipole approximation (green) shown as a function of the detectors spatial or temporal separation $s/\ell$.}\label{fig:wth}
\end{figure}

In Fig.~\ref{fig:wth} we plot the thermal Wightman function and its multipole expansion for an inverse temperature $\beta = 50\ell$ as a function of their separation when $\Delta t_{ij} = 0$ and $\Delta x_{ij} = s$, as well as when they are separated only in time, $\Delta t_{ij} = s$ and $\Delta x_{ij} = 0$. The vacuum Wightman function is displayed in blue for comparison. Same as in the case of the Minkowski vacuum, the multipole approximation becomes more accurate as $s/\ell$ increases. We picked $\beta = 50\ell$ so that $\ell \ll \beta$, ensuring that the multipole approximation is valid.

\subsection{Coherent state}\label{sec:coherent}

In this subsection we will consider the multipole expansion of a coherent state in Minkowski spacetime.   
A coherent state can be defined by a source $g_0\in C^\infty_0(\M)$ which defines a classical solution $\phi_0 = Eg_0$. 
Then a coherent state excitation of a state $\w$ is described by the mapping
$\w(\,\cdot \, ) \mapsto\omega_{g_0}(\,\cdot\,) = \w(e^{\ii \hphi(g_0)}~\cdot~e^{-\ii \hphi(g_0)})$. Here we will focus on a coherent state obtained by applying the displacement operator $e^{-\ii \hat{\phi}(g_0)}$ to the Minkowski vacuum, $\ket{g_0} =  e^{-\ii \hat{\phi}(g_0)}\ket{0}$. The corresponding two-point function will then be given by
\begin{equation} 
    W_{g_0}(\mf x_i,\mf x_j)
    =
    W_0(\mf x_i,\mf x_j)+\phi_0(\mf x_i)\phi_0(\mf x_j),
\end{equation}
and unlike the previous two examples, its one-point functions are non-vanishing, $\langle\hat{\phi}(f)\rangle_{g_0} = \phi_0(f)$.

Notice that, in principle, the protocol of Section~\ref{sec:extract} for extracting two-point functions of QFTs using local probes relies on the assumption that the field state is a \textit{zero mean} Gaussian state. Indeed, for a displaced Gaussian state, Eqs.~\eqref{eq:Hab simple} and~\eqref{eq:HC} would instead yield the second moments \mbox{$\langle\{\hat{\phi}(\Lambda_{\mf x_i}) - \langle \hat{\phi}(\Lambda_{\mf x_i})\rangle,\hat{\phi}(\Lambda_{\mf x_j}) - \langle \hat{\phi}(\Lambda_{\mf x_j})\rangle\}\rangle$}. In this sense, if applied to the coherent state $e^{-\ii \hat{\phi}(g_0)}\ket{0}$, the protocol of Section~\ref{sec:extract} would yield the vacuum two-point function, as the second moments are unchanged by the Weyl operator $e^{-\ii \hat{\phi}(g_0)}$. Nevertheless, we can still study its multipole expansion.

For an explicit example, we choose our source function, $g_0(\mf x)$, to be the spacetime Gaussian
\begin{equation}\label{eq:g0}
     g_0(\mf x) = \frac{e^{-\frac{t^2 + |\bm x|^2}{\delta^2}}}
    {(2\pi^2)^{3/2}\delta^3}
\end{equation}
centered at the origin. The normalization of Eq.~\eqref{eq:g0} ensures that the classical solution $\phi_0 = Eg_0$ has units of energy. In Appendix~\ref{app:coherent} we show that the corresponding classical field is
\begin{equation}\label{eq:coherent}
    \phi_0(\mf x)
    =
    \frac{e^{-\frac{(|\bm x|+t)^2}{4 \delta ^2}}-e^{-\frac{(|\bm x|-t)^2}{4 \delta ^2}}}{4 \sqrt{2} \pi  |\bm x|}.
\end{equation}
The classical solution above is composed of a positive-valued wave coming from past null infinity and a negative-valued wave propagating to future null infinity.

\definecolor{forestgreen}{RGB}{100,200,100}

\definecolor{forestgreen2}{RGB}{50,150,50}

\begin{figure}[h!]
    \centering
    \begin{tikzpicture}
        \node at (0,0) {\includegraphics[width=\linewidth]{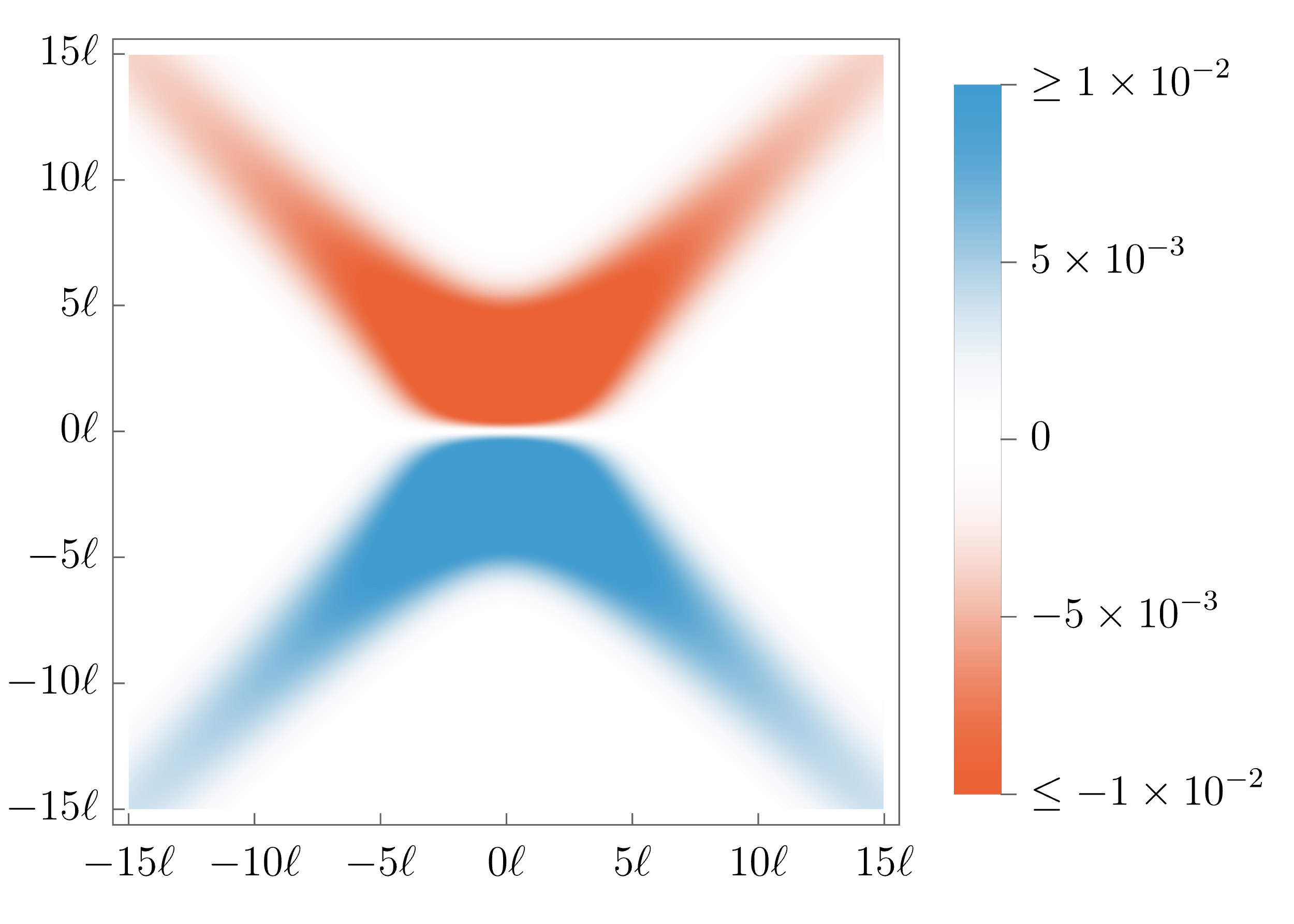}}; 
        \filldraw[black] (-1.9,-0.85) circle (1.5pt) ;
        \draw[black,thick,-] (-1.1,-0.85+0.07) -- (-1.1,-0.85-0.07); 
        \draw[black,thick,dashed,->] (-1.1,-0.85) -- (1.43,-0.85); 
        \draw[black,thick,-] (-1.9+0.07,0) -- (-1.9-0.07,0); 
        \draw[black,thick,dashed,->] (-1.9,0) -- (-1.9,2.5); 
        \node at (-1.9,-1.15) {$\mf x_i$};        
        \node[black] at (0.25,-1.15) {\footnotesize{$\Delta x_{ij} = s$}};
        \node[black,rotate=90] at (-2.15,1.25)  {\footnotesize{$\Delta t_{ij} = s$}};
    \end{tikzpicture}
    \caption{The classical solution $\phi_0(\mf x)$ associated with the coherent state $\ket{g_0}$ with $\delta = 3/2\ell$ in the $(t,x)$ plane. The black dot denotes the event $\mf x_i$ used in Fig.~\ref{fig:wcoh}, which is fixed, and the dashed lines denote the placements later chosen for $\mf x_j$.}
    \label{fig:coh}
\end{figure}
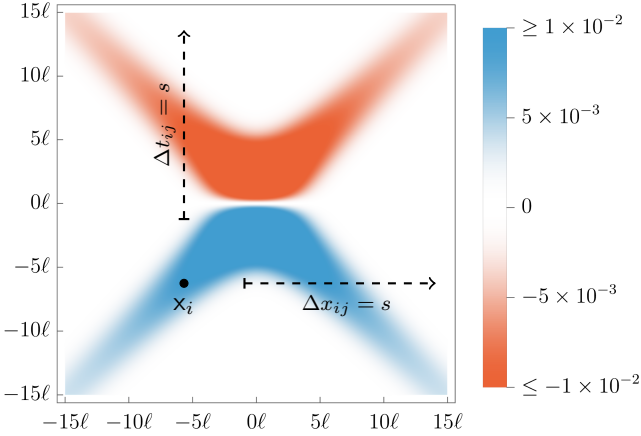

\begin{figure}[h!]
    \centering
    \includegraphics[width=\linewidth]{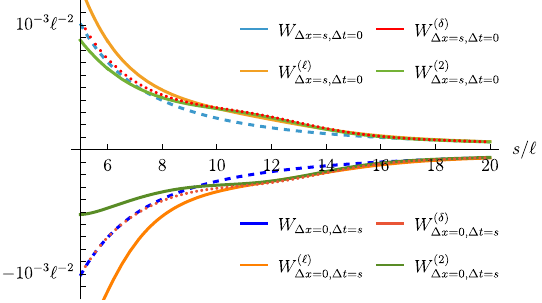}
    \caption{The two-point function for the Minkowski vacuum (blue), the coherent state of Eq.~\eqref{eq:coherent} with $\delta = \frac{3}{2}\ell$ (red), the smeared coherent state Wightman function (orange) and its multipole approximation (green). We fix $x_i = - 6\ell$, $t_i = 6\ell$ and vary $\mf x_j$ as a function of $s/\ell$ both when $x_j = -6\ell + s$ and $\Delta t_{ij} = 0$ and when $t_j = 6\ell-s$ and $\Delta x_{ij} = 0$.}
    \label{fig:wcoh}
\end{figure}


In Fig.~\ref{fig:coh} we plot the classical solution $\phi_0(\mf x)$ for $\delta = \frac{3}{2}\ell$ in the $(t,x)$ plane. We see that the difference is more prominent along the lightcone sourced by $g_0(\mf x)$, centered at $(0,\bm 0)$. In Fig.~\ref{fig:wcoh} we show the Wightman function $W_\varphi(\mf x_i,\mf x_j)$ (in red) and its multipole expansion (in green) when $\mf x_i$ is fixed at $(-6\ell, -6 \ell,0,0)$ and $\mf x_j$ is varied with either $\Delta t_{ij} = 0$, $\Delta x_{ij} = s$, or $\Delta t_{ij} = s$, $\Delta x_{ij} = 0$ for $\delta = 4\ell$ (see Fig.~\ref{fig:coh} where the dashed lines represent the positions of the event $\mf x_j$). The dashed blue lines in Fig.~\ref{fig:wcoh} display the vacuum Wightman function for comparison. Notice that for this choice of $\mf x_i$ and varying $\mf x_j$ along the lines depicted in Fig.~\ref{fig:coh}, $W_\varphi(\mf x_i,\mf x_j)$ has a slight peak at $\Delta t_{ij}=0$ and $\Delta x_{ij} \approx 12\ell$ and at $\Delta t_{ij}\approx 12\ell$ and $\Delta x_{ij}= 0$. These peaks happen when $\mf x_j$ overlaps with the lightcone of propagation of the classical solution. Same as in the previous examples, the approximation becomes more accurate as $s/\ell$ increases. We also see that when $\mf x_j$ is sufficiently distant from the propagation lightcone, the coherent correlation function approaches the vacuum two-point function.


\subsection{One particle state}

In this subsection we study the multipole expansion when the field is in a one particle wavepacket. Using the plane wave solutions to the massless Klein-Gordon equation
\begin{equation}
    u_{\bm k}(\mf x) = \frac{1}{\sqrt{2 |\bm k|}}\frac{e^{-\ii |\bm k| t + \bm k\cdot \bm x}}{(2\pi)^{3/2}},
\end{equation}
the field can be expanded in terms of creation and annihilation operators as
\begin{equation}
    \hat{\phi}(\mf x) = \int \dd^3 \bm k \left(u_{\bm k}(\mf x) \hat{a}_{\bm k}+u_{\bm k}^*(\mf x) \hat{a}_{\bm k}^\dagger\right),
\end{equation}
where $\mf k = (|\bm k|, \bm k)$ and the creation and annihilation operators satisfy the canonical commutation relations $[\hat{a}_{\bm k},\hat{a}_{\bm k'}^\dagger] = \delta^{(3)}(\bm k - \bm k')$. In this context the vacuum can be represented as a normalized state $\ket{0}$ such that $\hat{a}_{\bm k}\ket{0} = 0$ for all $\bm k$. A general one-particle wavepacket can then be written as
\begin{equation}
    \ket{\varphi} = \int \dd^3\bm k f(\bm{k}) \hat{a}_{\bm k}^\dagger \ket{0},
\end{equation}
where $f(\bm k)$ determines the contribution of each mode to the wave-packet. Notice that a one-particle wave packet is not a Gaussian state, and thus, the protocol of Section~\ref{sec:extract} does not apply---we focus on its multipole expansion.

For convenience, we will work with a wavepacket defined by
\begin{equation}\label{eq:f(k)}
    f(\bm k) =
    \frac{\delta^2}{\sqrt{4\pi}}
    \sqrt{2 |\bm{k}|} 
    e^{-\frac{\delta^2 \bm k^2}{2}},
\end{equation}
where $\delta$ is a constant with units of length. The Wightman function of the state $\ket{\varphi}$ is then given by
\begin{equation}\label{eq:tilde W}
    W_{\varphi}(\mf x,\mf x') =
    W_0(\mf x, \mf x') + F(\mf x)F(\mf x')^*+F(\mf x')F(\mf x)^*
\end{equation}
where
\begin{equation}
    F(\mf x)
    =
    \int \dd^3 \bm k ~
    u_{\bm k}(\mf x) f(\bm k)
\end{equation}
is the corresponding positive frequency solution determining the shape of the state in spacetime through its contribution to the two-point function. In particular, with the choice of $f(\bm k)$ of Eq.~\ref{eq:f(k)}, the one-particle wavepacket in localized around $\bm x = 0$ at $t = 0$ with spread $\delta$ in space and propagates away from the origin through the lightcone as $t$ increases. 

With the choice of Eq.~\eqref{eq:f(k)}, both $F(\mf x)$ and $W_\varphi(\mf x, \mf x')$ can be computed in closed form. In Appendix~\ref{app: One particle state} we show that
\begin{equation}
    F(\mf x) = \frac{1}{2|\bm x|}\left(H^{-}\!\!\left(\tfrac{|\bm x|-t}{\sqrt{2}\delta}\right)+H^{+}\!\!\left(\tfrac{|\bm x|+t}{\sqrt{2}\delta}\right)\right),
\end{equation}
where 
\begin{equation}
    H^\pm(v) = \frac{v e^{-v^2}}{\sqrt{2\pi}}\left(1 \pm \ii \text{erfi}(v)\right).
\end{equation}
Using the results of Section~\ref{sec:error estimation} we can then obtain the leading order multipole expansion of the Wightman function. However, we could not compute the smeared two-point functions $W_\varphi(\Lambda_{\mf x_i},\Lambda_{\mf x_j})$ in closed-form from these expressions.

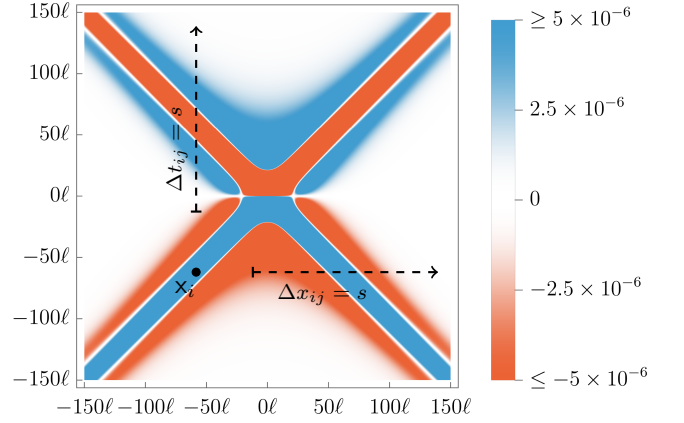
\begin{figure}[h!]
    \centering
    \begin{tikzpicture}
        \node at (0,0) {\includegraphics[width=\linewidth]{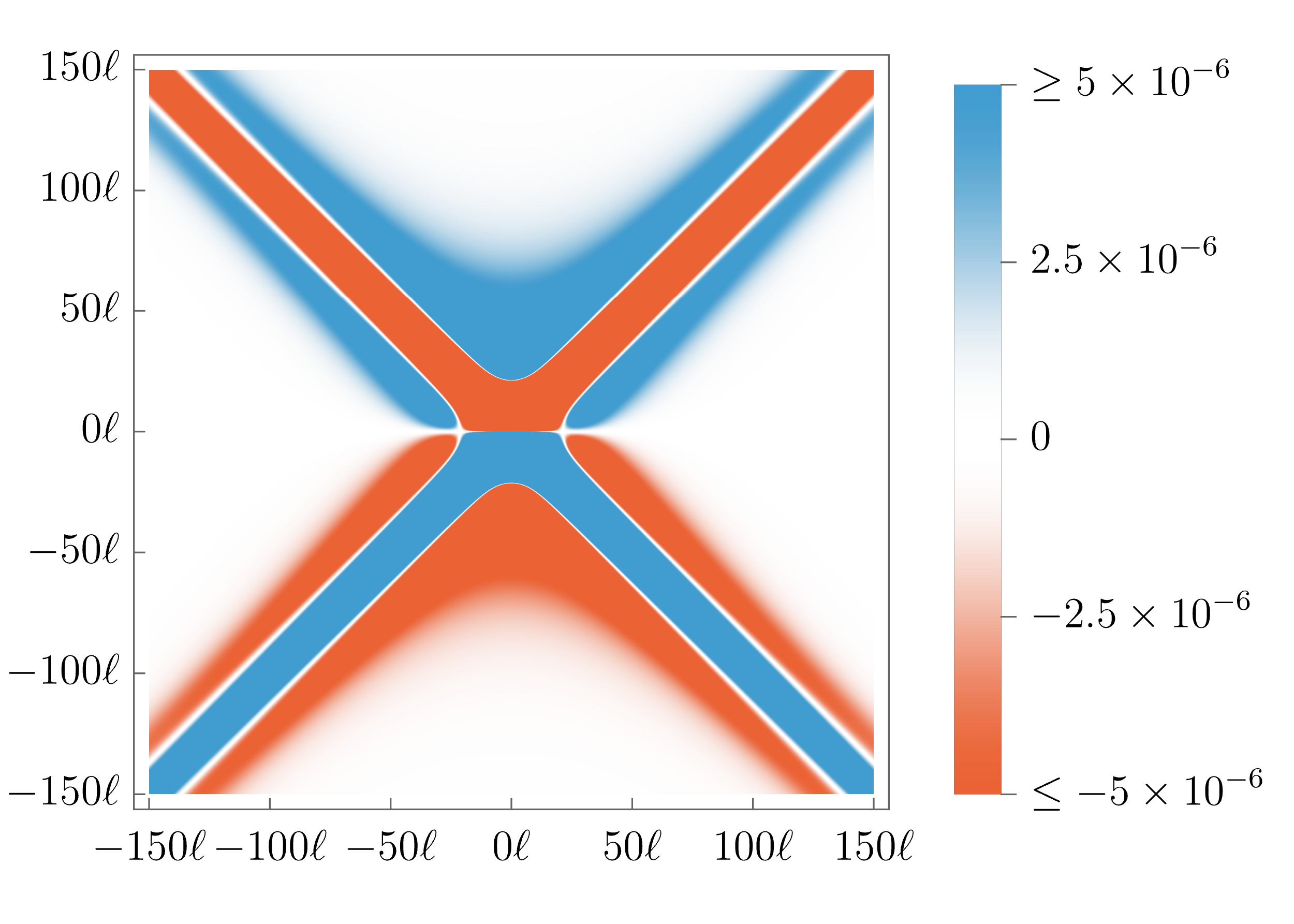}}; 
        \filldraw[black] (-1.85,-0.8) circle (1.5pt) ;
        \draw[black,thick,-] (-1.1,-0.8+0.07) -- (-1.1,-0.8-0.07); 
        \draw[black,thick,dashed,->] (-1.1,-0.8) -- (1.35,-0.8); 
        \draw[black,thick,-] (-1.85+0.07,0) -- (-1.85-0.07,0); 
        \draw[black,thick,dashed,->] (-1.85,0) -- (-1.85,2.45); 
        \node at (-1.99,-1.02) {$\mf x_i$};       
        \node[black] at (-0.18,-1.1) {\footnotesize{$\Delta x_{ij} = s$}};
        \node[black,rotate=90] at (-2.1,0.82)  {\footnotesize{$\Delta t_{ij} = s$}};
    \end{tikzpicture}
    \caption{The difference between the correlation function of the one-particle state for $\delta = 10\ell$ and the vacuum Wightman function in the $(t,x)$ plane. The black dot denotes the event $\mf x_i$ used in Fig.~\ref{fig:1 particle}, which is fixed, and the dashed lines denote the placements of $\mf x_j$.}\label{fig:ST-diagram}
\end{figure}

\begin{figure}[h!]
    \centering
    \includegraphics[width=\linewidth]{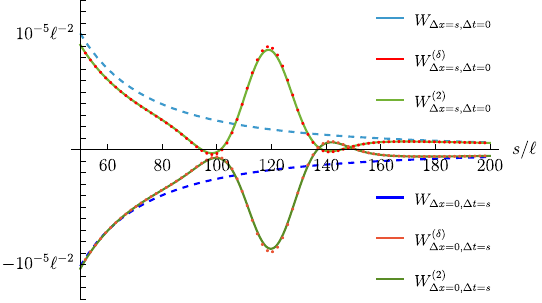}
    \caption{    
    The two point function for the one particle wavepacket~\eqref{eq:f(k)} with $\delta = 10\ell$ (red), its leading order multipole approximation (green) and the vacuum Wightman function (blue). We fix $x_i = - 60\ell$, $t_i = -60\ell$ and vary $\mf x_j$ as a function of $s/\ell$ both when $x_j = -60\ell + s$ and $\Delta t_{ij} = 0$ and when $t_j = -60\ell+s$ and $\Delta x_{ij} = 0$.}\label{fig:1 particle}
\end{figure}

In Fig.~\ref{fig:ST-diagram} we plot the difference of the particle and vacuum Wightman function, $W_{\varphi}(\mf x_i,\mf x) - W_0(\mf x_i,\mf x)$ for $\delta = 10\ell$, where $\mf x_i = (-60\ell,-60\ell,0,0)$, as a function of the spacetime position of the second event $\mf x = (t,x,0,0)$. We see that the difference is more prominent along the lightcone of the particle excitation, centered at $(0,\bm 0)$, which contains non-negligible information about the correlations within the wavepacket. This lightcone corresponds to the propagation of the state, which becomes localized around the origin with characteristic size $\delta$ at $t=0$. In Fig.~\ref{fig:1 particle} we show the Wightman function $W_\varphi(\mf x_i,\mf x_j)$ (in red) and its multipole expansion (in green) when again $\mf x_i = (-60\ell, -60 \ell,0,0)$, $\delta = 10\ell$, and either $\Delta t_{ij} = 0$, $\Delta x_{ij} = s$, or $\Delta t_{ij} = s$, $\Delta x_{ij} = 0$ (see Fig.~\ref{fig:ST-diagram} where the dashed lines represent the positions of the event $\mf x_j$). The dashed blue lines in Fig.~\ref{fig:1 particle} display the vacuum Wightman function for comparison. Notice that for this choice of $\mf x_i$ and varying $\mf x_j$ along the lines depicted in Fig.~\ref{fig:ST-diagram}, $W_\varphi(\mf x_i,\mf x_j)$ has a positive peak at $\Delta t_{ij}=0$ and $\Delta x_{ij} \approx 120\ell$ and a negative peak at $\Delta t_{ij}\approx 120\ell$ and $\Delta x_{ij}= 0$. These peaks happen when $\mf x_j$ overlaps with the lightcone of propagation of the wavepacket. Same as in the case of the Minkowski vacuum, we see that the approximation becomes more accurate as $s/\ell$ increases. We also see that when $\mf x_j$ is sufficiently distant from the propagation lightcone, the wavepacket correlation function approaches the vacuum two-point function.

\section{Conclusions}\label{sec:conclusions}

We developed a protocol for recovering a quantum field's two-point function between arbitrary spacetime regions. By considering a lattice of sequentially coupled gapless particle detectors, we identified the specific detectors' observables that encode the smeared field's two-point function after their interactions. When the probes couple to the field in finite regions of spacetime, the obtained two-point function is smeared against the detectors' interaction regions. We also quantified the difference between the smeared and point-wise two-point function through a multipole expansion, qualifying the regime where the protocol can be employed. 

Our non-perturbative results improve on previous two-point function measurement protocols~\cite{pipo,geometry} in three complementary ways. First, our results are valid in both the perturbative limit as well as the regime of strong coupling constants. Second, unlike~\cite{pipo,geometry}, our protocol also yields finite results even in the limit where the detectors' interactions become Dirac deltas centered at spacetime events, allowing one to recover the two-point function between arbitrarily small regions. Third, the gapless model we employ prevents the detectors' internal dynamics from hindering the recovery of the two-point function.


Our protocol outlines a clear path for directly accessing the two-point function of any field state in any globally hyperbolic spacetime. Taking into account that many states are fully characterized by their two-point function, access to this quantity is essential when determining which state can accurately model any given physical setup. Moreover, in~\cite{achim,achim2,geometry} it was shown that the short scale behaviour of the field's correlation function contains full information about the geometry of spacetime, allowing one to reconstruct the metric from the correlations in QFT. As a consequence, in the limit where the detectors' separation is sufficiently small, our protocol can also be used to measure distances and times.

Beyond its  applications to state reconstruction, the protocol developed here also highlights the operational viewpoint of QFT in curved spacetimes. Rather than relying on a preferred notion of particles or vacuum state, it shows that physically relevant information about a quantum field can be inferred directly from local measurements. In this context, our results provide a connection between the algebraic characterization of states in QFT and concrete measurement schemes~\cite{FewsterVerch,fewster2,jose}, outlining how field correlations can be accessed, approximated, and used to probe both matter and geometry.\\[3mm]

\acknowledgements

The authors thank Marcos Morote-Balboa for comments on the first draft of the manuscript. TRP is thankful for financial support from the Olle Engkvist Foundation (no.225-0062). Nordita is partially supported by Nordforsk.

\bibliography{references}

\appendix

\onecolumngrid

\section{Details for evolution}\label{app:evolution}

In this appendix we go over the details of for determining  the final state of the detectors.
Given an initial state for an N detector and field system, $\r_0=\r_{D,0}\otimes\r_{\phi,0}$, with the detectors in the state $\r_{D,0}$ and the field in the state $\r_{\phi,0}$, we wish to describe the composite state after the interactions have taken place. This evolution is given by application of the time evolution operator $\hat{U}$ defined in Eq.~\eqref{eq U definition}. We will rewrite this in terms of the Magnus expansion, keeping only the first two terms since  triple commutators of $\hat{h}_I(\mf x)$ vanish. We can then write
\begin{equation}
    \hat{U} = e^{\hat{\Theta}_1+\hat{\Theta}_2},
\end{equation}
where 
\begin{equation}
    \begin{aligned}
        \hat{\Theta}_1 
        &=
        -\ii \int\dd V
        \hat{h}_i(\mf x)=
        -\ii \int\dd V
        \lambda \sum_{i=1}^N \hat{\mu}_i \Lambda_i \hphi(\mf x) =
        -\ii \lambda
        \sum_{i=i}^N
        \hat{\mu}_i \hphi(\Lambda_i),
    \end{aligned}
\end{equation}
and 

\begin{align}
    \hat{\Theta}_2 
    &=
    \frac{-\ii }{2} \int\int \dd V'\dd V
    \theta(t-t')
    \comm{\hat{h}_i(\mf x)}{\hat{h}_i(\mf x')}
    \\
    &=
    \frac{- \lambda^2}{2}  \int\int \dd V'\dd V
    \theta(t-t')
    \comm{ \sum_{i=1}^N \hat{\mu}_i \Lambda_i \hphi(\mf x)}{\sum_{i=1}^N \hat{\mu}_i \Lambda_i \hphi(\mf x')}  
    \nonumber
    \\
    &=
    \frac{- \lambda^2}{2} \int\int \dd V'\dd V
    \theta(t-t')
    \comm{\hphi(\mf x)}{\hphi(\mf x')}
    \sum_{i,j=1}^N 
    \hat{\mu}_i \hat{\mu}_j \Lambda_i \Lambda_j ,\nonumber
\end{align}
with $\theta(t-t')$ being the Heviside step function.
By using $\theta(t-t')[{\hphi(\mf x)}{\hphi(\mf x')}]= \ii G_R(\mf x,\mf x')$ and $\Delta(\mf x, \mf x') = G_R(\mf x,\mf x')+G_R(\mf x',\mf x)$ we can rewrite $\hat{\Theta}_2$ as
\begin{align}
        \hat{\Theta}_2 
        &=
        \frac{-\ii \lambda^2}{2} \int\int \dd V'\dd V
        G_R(\mf x, \mf x')
        \sum_{i=1}^N
        \hat{\mu}_i^2 \Lambda_i \Lambda_i
        +
        G_R(\mf x, \mf x')
        \sum_{\substack{i,j =1 \\ i \neq j}}^N 
        \hat{\mu}_i \hat{\mu}_j \Lambda_i \Lambda_j  \\  
         &=
        \frac{-\ii \lambda^2}{2}
        \left(
        \sum_{i=1}^N
        \hat{\mu}_i^2 G_R(\Lambda_i,\Lambda_i)
        +
        \sum_{\substack{i,j =1 \\ i < j}}^N 
        \hat{\mu}_i \hat{\mu}_j \Delta(\Lambda_i,\Lambda_j) 
        \right).\nonumber
\end{align}
From this point on we will assume that $\hat{\mu}_i \propto \hat{\sigma}_x^{(i)}$, so that terms of the form $\hat{\mu}_i^2 G_R(\Lambda_i,\Lambda_i)$ correspond to global phases and can be neglected. The operators $\hat{\Theta}_1$ and $\hat{\Theta}_2$ also commute, so that
\begin{equation}
    \hat{U}  = \hat{U}_\phi\hat{U}_d = \exp\left(
        -\ii \lambda
        \sum_{i=1}^N
        \hat{\mu}_i \hphi(\Lambda_i)
        \right)
        \exp\left(
        \frac{-\ii \lambda^2}{2}
        \sum_{\substack{i,j =1 \\ i < j}}^N 
        \hat{\mu}_i \hat{\mu}_i \Delta(\Lambda_i,\Lambda_i) 
        \right).
\end{equation}
To get the final state of the field and detectors, $\r_f=\r_\tc{d}\otimes\r_\phi$, we need to act with the time evolution operator and its conjugate transpose on the initial state as
\begin{equation}
    \r_f = \hat{U}\r_0\hat{U}^\dagger.
\end{equation}
Since only $\hat{U}_\phi$ depends on the field and the final detector state is given by tracing out the field degrees of freedom, we may simplify the final detector state as
\begin{equation}
    \begin{aligned}
        \r_\tc{d} 
        &= 
        \Tr_\phi(\hat{U}\r_0\hat{U}^\dagger) \\
        &=
        \hat{U}_d \Tr_\phi (\hat{U}_\phi\r_0\hat{U}_\phi^\dagger)\hat{U}_d^\dagger.
    \end{aligned}
\end{equation}
We proceed by calculating the partial trace above
\begin{equation}\label{Appendixeq traceshit}
    \begin{aligned}
        \Tr_\phi (\hat{U}_\phi\r_0\hat{U}_\phi^\dagger)
        &= 
        \Tr\left(
        \exp\left(
        -\ii \lambda
        \sum_{i=1}^N
        \hat{\mu}_i \hphi(\Lambda_i)
        \right)
        ~
        \r_0
        ~
        \exp\left(
        \ii \lambda
        \sum_{i=1}^N
        \hat{\mu}_i \hphi(\Lambda_i)
        \right)
        \right).
    \end{aligned}
\end{equation}
We will expand the expression by inserting the identity in the $\hmu$ basis twice. Before this we will briefly make clear what bases are in use.  In the basis of eigenstates of $\hat{\sigma}_z$, $\{\ket{g},\ket{e}\}$, we can write in matrix form:
\begin{equation}
    \ket{g} = 
    \begin{pmatrix}
    1 \\ 0    
    \end{pmatrix}
    ~~~~~~~~~~~~
    \ket{e} = 
    \begin{pmatrix}
    0 \\ 1    
    \end{pmatrix},
\end{equation}
\begin{equation}
    \sm = 
    \begin{pmatrix}
    0&1 \\ 0&0    
    \end{pmatrix}
    ~~~~~~~~~~~~
    \sp = 
    \begin{pmatrix}
    0&0 \\ 1&0    
    \end{pmatrix},
\end{equation}
and
\begin{equation}
    \hmu
    =
    \begin{pmatrix}
        0&1\\1&0
    \end{pmatrix},
\end{equation}
with eigenvectors
\begin{equation}
    \ket{+} =
    \frac{1}{\sqrt{2}}
    \begin{pmatrix}
    1 \\ 1    
    \end{pmatrix},
    ~~~~~~~~~~~~
    \ket{-} =
    \frac{1}{\sqrt{2}}
    \begin{pmatrix}
    -1 \\ \,1    
    \end{pmatrix},
\end{equation}
with corresponding eigenvalues $\mu_+=1$ and $\mu_-=-1$. We now insert the identity in the basis of eigenvectors of $\hmu$ twice into Eq.~\eqref{Appendixeq traceshit}, giving 
\begin{align}
        \Tr_\phi (\hat{U}_\phi\r_0\hat{U}_\phi^\dagger)
        &= 
        \Tr_\phi
        \sum_{\substack{\mu_1,\mu_1'= \pm \\ ... \\\mu_N,\mu_N'= \pm}}
        \left(
        e^{
        -\ii \lambda
        \sum_{i=1}^N
        \mu_i \hphi(\Lambda_i)
        }
        ~
        \ket{\mu_1...\mu_N}\!\!
        \bra{\mu_1...\mu_N}
        \r_0
        \ket{\mu_1'...\mu_N'}\!\!
        \bra{\mu_1'...\mu_N'}
        ~
        e^{
        \ii \lambda
        \sum_{j=1}^N
        \mu_j' \hphi(\Lambda_j)
        }
        \right) \\
        &=
        \Tr_\phi
        \sum_{\substack{\mu_1,\mu_1'= \pm \\ ... \\ \mu_N,\mu_N'= \pm}}
        \left(
        e^{
        -\ii \lambda
        \sum_{i=1}^N
        \mu_i \hphi(\Lambda_i)
        }
        ~
        \bra{\mu_1...\mu_N}
        \r_0
        \ket{\mu_1'...\mu_N'}
        \ket{\mu_1...\mu_N}\!\!
        \bra{\mu_1'...\mu_N'}
        ~
        e^{
        \ii \lambda
        \sum_{j=1}^N
        \mu_j' \hphi(\Lambda_j)
        }
        \right) \nonumber\\
        &=\!\!\!\!\!
         \sum_{\substack{\mu_1,\mu_1'= \pm \\ ... \\ \mu_N,\mu_N'= \pm}}
         \left(
        \w \left(
        e^{
        -\ii \lambda
        \sum_{i=1}^N
        \mu_i \hphi(\Lambda_i)
        }
        e^{
        \ii \lambda
        \sum_{j=1}^N
        \mu_j' \hphi(\Lambda_j)
        }
        \right)
        ~
        \bra{\mu_1...\mu_N}
        \r_{\tc{d},0}
        \ket{\mu_1'...\mu_N'}
        \ket{\mu_1...\mu_N}\!\!
        \bra{\mu_1'...\mu_N'}
        \right)\nonumber \\
        &=\!\!\!\!\!
         \sum_{\substack{\mu_1,\mu_1'= \pm \\ ... \\ \mu_N,\mu_N'= \pm}}
         \!\!\!\!\!\!
        e^{\frac{\ii\lambda^2}{2}
        E\big(\sum_{i=1}^N
        \mu_i \hphi(\Lambda_i)
        ,\sum_{j=1}^N
        \mu_j' \hphi(\Lambda_j)\big)
        -\frac{\lambda^2}{2}
        ||\sum_{i=1}^N
        (\mu_i-\mu_i') \Lambda_i||^2}
        \bra{\mu_1...\mu_N}
        \r_{\tc{d},0}
        \ket{\mu_1'...\mu_N'}
        \ket{\mu_1...\mu_N}\!\!
        \bra{\mu_1'...\mu_N'} \nonumber,
\end{align}
where we have used the relation
\begin{equation}
    \w\left(
    e^{\ii\lambda\hphi(f)}e^{\ii\lambda\hphi(g)}
    \right)
    =
    e^{\frac{\ii\lambda^2}{2}E(f,g)-\frac{\lambda^2}{2}W(f+g,f+g)}
\end{equation}
in the last equality, as well as the short-hand notation $||f||^2 = W(f,f)$. To get the full final final state of the detectors state we also need to apply the unitary $\hat{U}_d$ and $\hat{U}_d^\dagger$, yielding the expression
\begin{align}
        \r_\tc{d} 
        &= 
        \hat{U}_d\Tr_\phi(\hat{U}_{\hphi} \r_0 \hat{U}_{\hphi}^\dagger)\hat{U}_d^\dagger   \\ 
        &= \sum_{\substack{\mu_1,\mu_1'= \pm \\ \dots \nonumber
        \\ 
        \mu_N,\mu_N'= \pm}}
        \exp\left(
        \frac{\ii \lambda}{2}
        \sum_{\substack{i,j=1 \\ i<j}}^N
        (\hmu_i'\hmu_j'-\hmu_i\hmu_j)\Delta(\Lambda_i,\Lambda_j)
        \right)
        \exp\left(
        \frac{\ii\lambda^2}{2}
        E\Big(\sum_{i=1}^N
        \mu_i' \hphi(\Lambda_i)
        ,\sum_{j=1}^N
        \mu_j \hphi(\Lambda_j)\Big)
        \right) \nonumber\\
        &~~~~~~~~~~~~~~~~
        \times\exp\left(
        -\frac{\lambda^2}{2}
        ||\sum_{i=1}^N
        (\mu_i-\mu_i') \Lambda_i||^2
        \right)
        ~
        \bra{\mu_1\dots\mu_N}
        \r_{\tc{d},0}
        \ket{\mu_1'\dots\mu_N'}
        \ket{\mu_1\dots\mu_N}
        \bra{\mu_1'\dots\mu_N'}\nonumber.
\end{align}
We will assume that the detectors all start in the ground state, that is $\r_{\tc{d},0} = \ket{g_{1}\dots g_{N}}\bra{g_{1}\dots g_{N}}$, we recall that the eigenvectors to $\hmu$ are 
\begin{equation}
    \ket{+} = \frac{1}{\sqrt{2}}
    \begin{pmatrix}
        1 \\ 1
    \end{pmatrix}
    =
    \frac{1}{\sqrt{2}}(\ket{g}+\ket{e}),
\end{equation}
\begin{equation}
    \ket{-} = \frac{1}{\sqrt{2}}
    \begin{pmatrix}
        1 \\ -1
    \end{pmatrix}
    =
    \frac{1}{\sqrt{2}}(\ket{g}-\ket{e}),
\end{equation}
which allows us to rewrite
\begin{align}
        &\r_{\tc{d},0} = \sum_{\substack{\mu_1,\mu_1'= \pm \\ \dots \\ \mu_N,\mu_N'= \pm}}
        \bra{\mu_1\dots\mu_N}
        \r_{\tc{d},0}
        \ket{\mu_1'\dots\mu_N'}
        \ket{\mu_1\dots\mu_N}
        \bra{\mu_1'\dots\mu_N'}
        =
        \sum_{\substack{\mu_1,\mu_1'= \pm \\ \dots \\ \mu_N,\mu_N'= \pm}}
        \frac{1}{2^N}
        \ket{\mu_1\dots\mu_N}
        \bra{\mu_1'\dots\mu_N'}\nonumber.
\end{align}
The identity above gives the final detector state
\begin{align}
        \r_\tc{d} 
        &= \sum_{\substack{\mu_1,\mu_1'= \pm \\ \dots \\ \mu_N,\mu_N'= \pm}}
        \exp\left(
        \frac{\ii \lambda}{2}
        \sum_{\substack{i,j=1 \\ i<j}}^N
        (\mu_i'\mu_j'-\mu_i\mu_j)\Delta(\Lambda_i,\Lambda_j)
        \right)
        \exp\left(
        \frac{\ii\lambda^2}{2}
        E\Big(\sum_{i=1}^N
        \mu_i' \hphi(\Lambda_i)
        ,\sum_{j=1}^N
        \mu_j \hphi(\Lambda_j)\Big)
        \right) \\
        &~~~~~~~~~~~~~~~~
        \exp\left(
        -\frac{\lambda^2}{2}
        ||\sum_{i=1}^N
        (\mu_i-\mu_i') \Lambda_i||^2
        \right)
        ~
        \frac{1}{2^N}
        \ket{\mu_1\dots\mu_N}
        \bra{\mu_1'\dots\mu_N'}\nonumber.
\end{align}
To simplify notation we define $\bm \mu = (\mu_1,...,\mu_N)$. With this notation we write the final detector state as 
\begin{align}
        \r_\tc{d} 
        &= \frac{1}{2^N}\sum_{\bm \mu,\bm \mu'}
        \exp\left(
        \frac{\ii \lambda^2}{2}
        \sum_{\substack{i,j=1 \\ i<j}}^N
        (\mu_i'\mu_j'-\mu_i\mu_j)\Delta(\Lambda_i,\Lambda_j)
        \right)
        \exp\left(
        \frac{\ii\lambda^2}{2}\sum_{i=1}^N \sum_{j=1}^N\mu_i'\mu_j 
        E(
        \Lambda_i
        ,
        \Lambda_j)
        \right) \\
        &~~~~~~~~~~~~~~~~
        \times\exp\left(
        -\frac{\lambda^2}{2}
       \sum_{i=1}^N
        \sum_{j=1}^N
        (\mu_i-\mu_i')(\mu_j-\mu_j')W( \Lambda_i, \Lambda_j)
        \right)
        ~
        \ket{\bm \mu}\!\!
        \bra{\bm \mu'}\nonumber.
\end{align}

\section{Details for expected values}\label{app:ev}
In this appendix we compute the nonzero expected values of the form $\ev{\s^{(i)}_\alpha\s^{(j)}_\beta}_{\r_\tc{d}}$. To calculate these quantities we first observe that
\begin{align}\label{ev non compressed}
        \langle\s^{(i)}_\alpha\s^{(j)}_\beta\rangle_{\r_\tc{d}}
        &=
        \Tr\left(
        \s^{(i)}_\alpha\s^{(j)}_\beta \r_\tc{d} 
        \right) \\
        &= \sum_{\substack{\mu_1,\mu_1'= \pm \\ \dots \\ \mu_N,\mu_N'= \pm}}
        \exp\left(
        \frac{\ii \lambda^2}{2}
        \sum_{\substack{k,l=1 \\ \ k<l}}^N
        (\mu_k'\mu_l'-\hmu_k\hmu_l)\Delta(\Lambda_k,\Lambda_l)
        \right)
        \exp\left(
        \frac{\ii\lambda^2}{2}
        E\Big(\sum_{k=1}^N
        \mu_k' \hphi(\Lambda_k)
        ,\sum_{l=1}^N
        \mu_l \hphi(\Lambda_l)\Big)
        \right) \nonumber\\
        &~~~~~~~~~~~~~~~~~~
        \exp\left(
        -\frac{\lambda^2}{2}
        ||\sum_{k=1}^N
        (\mu_k-\mu_k') \Lambda_k||^2
        \right)
        ~
        \frac{1}{2^N}
        \bra{\mu_i'\mu_j'}
        \s^{(i)}_\alpha\s^{(j)}_\beta
        \ket{\mu_i\mu_j}
        \Tilde{\delta}(i,j),\nonumber
\end{align}
with $\Tilde{\delta}(i,j) = 
\prod_{\substack{k=1 \\ k\neq i,j}}
\delta_{\mu_k,\mu_k'}$. 
We may simplify Eq.~\eqref{ev non compressed} by further by manipulating the individual exponential terms. The presence of $\Tilde{\delta}(i,j)$ in the first exponential factor, as well as symmetry of $\Delta$ allows us to recast it as
\begin{align}
        \frac{\ii}{2}
        \sum_{\substack{k,l=1 \\ k<l}}^N
        (\mu_k'\mu_l'-\hmu_k\hmu_l)\Delta_{kl}
        &=
        \frac{\ii}{2}\Bigg(
        (\mu_k'-\mu_k)
        \sum_{\substack{k=1 \\k\neq i,j}}^N
        \mu_k\Delta_{ki}
        +
        (\mu_j'-\mu_j)
        \sum_{\substack{k=1 \\k\neq i,j}}^N
        \mu_k\Delta_{kj}
        +
        (\mu_i'\mu_j'-\mu_i\mu_j)\Delta_{ij}
        \Bigg)
        \\
        &=
        \frac{\ii}{2}\Bigg(
        (\mu_i'-\mu_i)
        \tilde{\Delta}_i
        +
        (\mu_j'-\mu_j)
        \tilde{\Delta}_j
        +
        (\mu_i'\mu_j'-\mu_i\mu_j)\Delta_{ij}
        \Bigg)\nonumber,
\end{align}
since terms will be zero if $\mu_i$ or $\mu_j$ are not in the term. We introduced short the shorthands $\lambda^2\Delta(\Lambda_i,\Lambda_j) = \Delta_{ij}$ and $\sum_{\substack{k=1 \\ k \neq i,j}}^N \mu_k\Delta_{ki}=\tilde{\Delta}_i$. We may also simplify the second exponential in Eq.~\eqref{ev non compressed}. We start by expanding the sums inside $E$ and simplify by using its bilinearity
\begin{align}
        \frac{\ii\lambda^2}{2}
        E&\Big(\sum_{k=1}^N
        \mu_k' \hphi(\Lambda_k)
        ,\sum_{l=1}^N
        \mu_l \hphi(\Lambda_l)\Big)\\
        &=
        \frac{\ii\lambda^2}{2}
        E
        \Bigg(
        \mu_i' \hphi(\Lambda_i)
        +
        \mu_j' \hphi(\Lambda_j)
        +
        \sum_{\substack{k=1\\ k \neq i,j}}^N
        \mu_k' \hphi(\Lambda_k),
        \mu_i \hphi(\Lambda_i)
        +
        \mu_j \hphi(\Lambda_j)
        +
        \sum_{\substack{k=1\\ k \neq i,j}}^N
        \mu_k \hphi(\Lambda_k)
        \Bigg)
         \nonumber\\
        &=
        \frac{\ii}{2}\Bigg(
        \mu_i'\mu_j E_{ij}
        +
        \mu_i'
        \sum_{\substack{k=1\\k\neq i,j}}
        \mu_k E_{ik}
        +
        \mu_j'\mu_i E_{ji}
        +
        \mu_j'
        \sum_{\substack{k=1\\k\neq i,j}}
        \mu_k E_{jk}\nonumber
        +
        \mu_i
        \sum_{\substack{k=1\\k\neq i,j}}
        \mu_k' E_{ki}
        +
        \mu_j
        \sum_{\substack{k=1\\k\neq i,j}}
        \mu_k' E_{kj}
        +
        \sum_{\substack{k,l=1\\k,l\neq i,j\\k\neq l}}
        \mu_k\mu_l' E_{kl}
        \Bigg),\nonumber
\end{align}
where we defined $E_{ij}=\lambda^2  E(\hphi(\Lambda_i),\hphi(\Lambda_j))$. Notice that the last sum is always zero, since the sum contains an $E_{ji}$ term for each $E_{ij}$ term and $E$ is antisymmetric. We continue to use the antisymmetry of $E$ and take $\Tilde{\delta}(i,j)$ into account, to get
\begin{align}
        \frac{\ii\lambda^2}{2}
        E\Big(\sum_{k=1}^N
        \mu_k' \hphi(\Lambda_k)
        ,\sum_{l=1}^N
        \mu_l \hphi(\Lambda_l)\Big)
        &=
        \frac{\ii}{2}\Bigg(
        (\mu_i'\mu_j - \mu_j'\mu_i)
        E_{ij}
        +
        (\mu_i'-\mu_i)
        \sum_{\substack{k=1\\k\neq i,j}}
        E_{ik}
        +
        (\mu_j'-\mu_j)
        \sum_{\substack{k=1\\k\neq i,j}}
        E_{jk} 
        \Bigg)\\
        &=
        \frac{\ii}{2}
        \Bigg(
        (\mu_i'\mu_j - \mu_j'\mu_i)
        E_{ij}
        +
        (\mu_i'-\mu_i)
        \tilde{E}_i(\bm \m)
        +
        (\mu_j'-\mu_j)
        \tilde{E}_j(\bm \m)
        \Bigg),\nonumber
\end{align}
where we introduced $\tilde{E}_i(\bm \m) =\sum_{\substack{k=1\\ k\neq ij}}\m_kE_{ik} $  above. Analogously, we define $\tilde{\Delta}_i(\bm \m) =\sum_{\substack{k=1\\ k\neq ij}}\m_k\Delta_{ik}$.
Let us now simplify the third and last exponential factor in Eq.~\eqref{ev non compressed}. Incorporating $\tilde{\delta}(i,j)$ we see that
\begin{equation}
    -\frac{\lambda^2}{2}
    ||\sum_{k=1}^\N(\mu_k-\mu_k')\Lambda_k||^2 
    =
    -\frac{\lambda^2}{2}
    ||
    (\mu_i-\mu_i')\Lambda_i
    +
    (\mu_j-\mu_j')\Lambda_j
    ||^2.
\end{equation}
Taking the simplifications above into account we may write Eq.~\eqref{ev non compressed} as 
\begin{align}
        \evv{\s^{(i)}_\alpha\s^{(j)}_\beta}_{\r_\tc{d}}
        &=
        \sum_{\substack{\mu_1,\mu_1'= \pm \\ \dots \\ \mu_N,\mu_N'= \pm}}
        \frac{1}{2^N}
        \bra{\mu_i'\mu_j'}
        \s^{(i)}_\alpha\s^{(j)}_\beta
        \ket{\mu_i\mu_j}
        ~
        \exp\left(
        -\frac{\lambda^2}{2}
        ||
        (\mu_i-\mu_i')\Lambda_i
        +
        (\mu_j-\mu_j')\Lambda_j
        ||^2
        \right)
        \\
        &
        ~~~~~~~~~~~~
        \times
        \exp
        \left(
        \frac{\ii}{2}\Bigg(
        (\mu_i'\mu_j'-\mu_i\mu_j)\Delta_{ij}
        +
        (\mu_i'-\mu_i)
        \tilde{\Delta}_i(\bm \m)
        +
        (\mu_j'-\mu_j)
        \tilde{\Delta}_j(\bm \m)
        \right)
        \Bigg)\nonumber
        \\
        &
        ~~~~~~~~~~~~
        \times
        \exp
        \left(
        \frac{\ii}{2}\Bigg(
        (\mu_i'\mu_j - \mu_j'\mu_i)
        E_{ij}
        +
        (\mu_i'-\mu_i)
        \tilde{E}_i(\bm \m)
        +
        (\mu_j'-\mu_j)
        \tilde{E}_j(\bm \m)
        \right)
        \Bigg).\nonumber
\end{align}\\
To lighten notation even further, we note that $\Delta_{ij} = G_{ij}+G_{ji}$ and $E_{ij}=G_{ij}-G_{ji}$ with $G_{ij}$ being the retarded propagator from $j$ to $i$, we see that $\tilde{\Delta}_i(\bm \m)+\tilde{E}_i(\bm \m) = 
2\sum_{\substack{k=1\\ k\neq i,j}}^N \mu_k G_{ik}$. We then define 
$\tilde{G}_i(\bm \m) \equiv \frac{1}{2}(
\tilde{\Delta}_i(\bm \m)+\tilde{E}_i(\bm \m))$. 
With this definition we may write the expected value as
\begin{equation}
    \begin{aligned}
        \evv{\s^{(i)}_\alpha\s^{(j)}_\beta}_{\r_\tc{d}}
        &=
        \sum_{\bm \mu, \bm \mu'}
        \frac{1}{2^N}
        \bra{\mu_i'\mu_j'}
        \s^{(i)}_\alpha\s^{(j)}_\beta
        \ket{\mu_i\mu_j}
        ~
        \exp\left(
        -\frac{\lambda^2}{2}
        ||
        (\mu_i-\mu_i')\Lambda_i
        +
        (\mu_j-\mu_j')\Lambda_j
        ||^2
        \right)
        \\
        &
        ~~~~~~~~~~~~
        \times
        \exp
        \left(
        \ii
        (\mu_i'-\mu_i)
        \tilde{G}_i(\bm \m)
        +
        \ii
        (\mu_j'-\mu_j)
        \tilde{G}_j(\bm \m)
        +
        \frac{\ii}{2}
        (\mu_i'\mu_j'-\mu_i\mu_j)\Delta_{ij}
        +
        \frac{\ii}{2}
        (\mu_i'\mu_j - \mu_j'\mu_i)
        E_{ij}
        \right) .
    \end{aligned}
\end{equation}
Finally, we will expand the first exponential and use the fact that the Hadamard function is bilinear and symmetric while the causal propagator is bilinear and antisymmetric to write
\begin{align}
        -\frac{\lambda^2}{2}
        ||(\m_i-\m_i')\Lambda_i+(\m_j-\m_j')\Lambda_j||^2
        &=
        -\frac{\lambda^2}{2}
        W\Big((\m_i-\m_i')\Lambda_i+(\m_j-\m_j')\Lambda_j~,~(\m_i-\m_i')\Lambda_i+(\m_j-\m_j')\Lambda_j\Big)
        \\
        &=
        -\frac{\lambda^2}{2}
        \Bigg[
        W\Big((\m_i-\m_i')\Lambda_i,(\m_i-\m_i')\Lambda_i\Big)
        +
        W\Big((\m_i-\m_i')\Lambda_i,(\m_j-\m_j')\Lambda_j\Big)
        \nonumber
        \\
        &
        ~~~~~~~~+
        W\Big((\m_j-\m_j')\Lambda_j,(\m_i-\m_i')\Lambda_i\Big)
        +
        W\Big((\m_j-\m_j')\Lambda_j,(\m_j-\m_j')\Lambda_j\Big)
        \Bigg]\nonumber
        \\
        &=
        -\frac{1}{2}
        \Bigg[
        (\m_i-\m_i')^2 W_{ii}
        +
        \frac{1}{2}(\m_i-\m_i')(\m_j-\m_j')H_{ij}\nonumber
        \\
        &
        ~~~~~~~+
        \frac{1}{2}(\m_j-\m_j')(\m_i-\m_i')H_{ji}
        +
        (\m_j-\m_j')^2 W_{jj}
        \Bigg]\nonumber
        \\
        &=
        -\frac{1}{2}
        \Bigg[
        (\m_i-\m_i')^2 W_{ii}
        +
        (\m_i-\m_i')(\m_j-\m_j')H_{ij}
        +
        (\m_j-\m_j')^2 W_{jj}
        \Bigg]\nonumber.
\end{align}
This allows us to recast the relevant expected values as
\begin{align}\label{appeq:general ev useful}
        \evv{\s^{(i)}_\alpha\s^{(j)}_\beta}_{\r_\tc{d}}
        &=
        \sum_{\bm \mu, \bm \mu'}
        \frac{1}{2^N}
        \bra{\mu_i'\mu_j'}
        \s^{(i)}_\alpha\s^{(j)}_\beta
        \ket{\mu_i\mu_j}
        ~
        \exp\left(
        -\frac{1}{2}
        \Big(
        (\m_i-\m_i')^2 W_{ii}
        +
        (\m_i-\m_i')(\m_j-\m_j')H_{ij}
        +
        (\m_j-\m_j')^2 W_{jj}
        \Big)
        \right)\nonumber
        \\
        &
        ~~~~~~~~~~~~
        \times
        \exp
        \left(
        \ii
        (\mu_i'-\mu_i)
        \tilde{G}_i(\bm \m)
        +
        \ii
        (\mu_j'-\mu_j)
        \tilde{G}_j(\bm \m)
        +
        \frac{\ii}{2}
        (\mu_i'\mu_j'-\mu_i\mu_j)\Delta_{ij}
        +
        \frac{\ii}{2}
        (\mu_i'\mu_j - \mu_j'\mu_i)
        E_{ij}
        \right) .
\end{align}
Throughout the next subsections we will use the expression above to compute the expected values for different combinations of $\alpha,\beta \in\{ 0,x,y,z\}$.

\subsection{Computing $\ev{\s_z^{(i)}\s_z^{(j)}}$}\label{app:evzz}
We start by computing $\langle{\s_z^{(i)}\s_z^{(j)}}\rangle$.
The first step is to evaluate the matrix elements $\bra{\mu_i'\mu_j'} \s_z^{(i)}\s_z^{(j)} \ket{\mu_i\mu_j}$ for $\ket{\mu_i\mu_j}\in\{\ket{++},\ket{+-}.\ket{-+},\ket{--}\}$. In this basis within the $ij$ subspace, we have the following matrix representation
\begin{equation}\label{appeq:double sigmazz}
    \s_z^{(i)}\s_z^{(j)}
    =
    \begin{pmatrix}
        0 & 0 & 0 & 1 \\
        0 & 0 & 1 & 0 \\
        0 & 1 & 0 & 0 \\
        1 & 0 & 0 & 0 \\
    \end{pmatrix}.
\end{equation}
Only the non-zero matrix elements will contribute in the final sum, so we will only have four non-trivial terms. In the table below we present the values of $\mu_i',\mu_j',\mu_i$ and  $\mu_j$ corresponding to non-zero values of $\bra{\mu_i'\mu_j'} \s_z^{(i)}\s_z^{(j)} \ket{\mu_i\mu_j}$:
\begin{center}
    \begin{tabular}{|c c c c | c|} 
         \hline
         $\mu_i'$ & $\mu_j'$ 
         & $\mu_i$ & $\mu_j$
         & $\bra{\mu_i'\mu_j'} \s_z^{(i)}\s_z^{(j)} \ket{\mu_i\mu_j}$\\ [0.5ex] 
         \hline\hline
          + & + & - & - & 1\\ 
         \hline
          + & - & - & + & 1\\
         \hline
          - & + & + & - & 1\\
         \hline
          - & - & + & + & 1\\ 
         \hline
    \end{tabular}.
\end{center}
Using the values of the table above, we can write the non-zero terms in the sum of Eq.~\eqref{appeq:general ev useful} one by one:
\begin{align}\label{evzz first}
        \evv{\s_z^{(i)}\s_z^{(j)}}
        &=
        \sum_{\substack{\mu_1,\mu_1'= \pm \\ \dots \\ \mu_N,\mu_N'= \pm}}
        \Bigg[
        \frac{1}{2^N}
        e^{\left(
        -\frac{\lambda^2}{2}
        ||
        -2\Lambda_i
        -2\Lambda_j
        ||^2
        \right)
        }
        \exp
        \left(
        2\ii
        \tilde{G}_i
        +
        2\ii
        \tilde{G}_j
        \right)\\
        &~~~~~~~~~~~~~~
        +
        \frac{1}{2^N}
        e^{\left(
        -\frac{\lambda^2}{2}
        ||
        -2\Lambda_i
        +2\Lambda_j
        ||^2
        \right)
        }
        \exp
        \left(
        2\ii
        \tilde{G}_i
        -
        2\ii
        \tilde{G}_j
        \right)\nonumber\\
        &~~~~~~~~~~~~~~
        +
        \frac{1}{2^N}
        e^{\left(
        -\frac{\lambda^2}{2}
        ||
        +2\Lambda_i
        -2\Lambda_j
        ||^2
        \right)
        }
        \exp
        \left(
        -2\ii
        \tilde{G}_i
        +
        2\ii
        \tilde{G}_j
        \right)\nonumber\\
        &~~~~~~~~~~~~~~
        +
        \frac{1}{2^N}
        e^{\left(
        -\frac{\lambda^2}{2}
        ||
        +2\Lambda_i
        +2\Lambda_j
        ||^2
        \right)
        }
        \exp
        \left(
        -2\ii
        \tilde{G}_i
        -
        2\ii
        \tilde{G}_j
        \right)
        \Bigg].\nonumber
\end{align}
The first exponentials can be rewritten as
\begin{align}
        -\frac{\lambda^2}{2}
        ||\pm2\Lambda_i\pm2\Lambda_j||^2
        &=
        -\frac{\lambda^2}{2}
        W(\pm2\Lambda_i\pm2\Lambda_j,
        \pm2\Lambda_i\pm2\Lambda_j) \\
        &=
        -\frac{\lambda^2}{2}
        W(\pm2(\Lambda_i+\Lambda_j),
        \pm2(\Lambda_i+\Lambda_j)) \nonumber\\
        &=
        -\frac{\lambda^2}{2}
        (\pm2)^2 W(\Lambda_i+\Lambda_j,
        \Lambda_i+\Lambda_j) \nonumber\\
        &=
        -2\lambda^2
        W(\Lambda_i+\Lambda_j,
        \Lambda_i+\Lambda_j)\nonumber
\end{align}
and
\begin{align}
        -\frac{\lambda^2}{2}
        ||\pm 2\Lambda_i\mp2\Lambda_j||^2
        &=
        -\frac{\lambda^2}{2}
        W(\pm2\Lambda_i\mp2\Lambda_j,
        \pm2\Lambda_i\mp2\Lambda_j) \\
        &=
        -\frac{\lambda^2}{2}
        W(\pm2(\Lambda_i-\Lambda_j),
        \pm2(\Lambda_i-\Lambda_j)) \nonumber\\
        &=
        -\frac{\lambda^2}{2}
        (\pm2)^2 W(\Lambda_i-\Lambda_j,
        \Lambda_i-\Lambda_j) \nonumber\\
        &=
        -2\lambda^2
        W(\Lambda_i-\Lambda_j,
        \Lambda_i-\Lambda_j),\nonumber
\end{align}
so that
\begin{align}\label{evzz before pull out}
        \evv{\s_z^{(i)}\s_z^{(j)}}
        &=
        \frac{1}{2^N}
        \sum_{\substack{\mu_1,\mu_1'= \pm \\ \dots \\ \mu_N,\mu_N'= \pm}}
        \Bigg[
        e^{-2W(\Lambda_i+\Lambda_j,
        \Lambda_i+\Lambda_j)
        }
        \exp
        \left(
        2\ii
        \tilde{G}_i
        +
        2\ii
        \tilde{G}_j
        \right) \\
        &~~~~~~~~~~~~~~~~~~~ 
        +
        e^{
        -2W(\Lambda_i-\Lambda_j,
        \Lambda_i-\Lambda_j)
        }
        \exp
        \left(
        2\ii
        \tilde{G}_i
        -
        2\ii
        \tilde{G}_j
        \right)\nonumber\\
        &~~~~~~~~~~~~~~~~~~~
        +
        e^{
        -2W(\Lambda_i-\Lambda_j,
        \Lambda_i-\Lambda_j)
        }
        \exp
        \left(
        -2\ii
        \tilde{G}_i
        +
        2\ii
        \tilde{G}_j
        \right)\nonumber\\
        &~~~~~~~~~~~~~~~~~~~
        +
        e^{
        -2W(\Lambda_i+\Lambda_j,
        \Lambda_i+\Lambda_j)
        }
        \exp
        \left(
        -2\ii
        \tilde{G}_i
        -
        2\ii
        \tilde{G}_j
        \right) 
        \Bigg]\nonumber
        \\
        & \nonumber\\
        &
        =
        \frac{1}{2^N}
        \sum_{\substack{\mu_1,\mu_1'= \pm \\ \dots \\ \mu_N,\mu_N'= \pm}}
        \Bigg[
        e^{-2W(\Lambda_i+\Lambda_j,
        \Lambda_i+\Lambda_j)
        }
        \left(
        \exp
        \left(
        2\ii
        \tilde{G}_i
        +
        2\ii
        \tilde{G}_j
        \right)
        +
        \exp
        \left(
        -2\ii
        \tilde{G}_i
        -
        2\ii
        \tilde{G}_j
        \right) 
        \right)\nonumber
        \\
        &~~~~~~~~~~~~~~~~~~~
        +
        e^{
        -2W(\Lambda_i-\Lambda_j,
        \Lambda_i-\Lambda_j)
        }
        \left(
        \exp
        \left(
        2\ii
        \tilde{G}_i
        -
        2\ii
        \tilde{G}_j
        \right)
        +
        \exp
        \left(
        -2\ii
        \tilde{G}_i
        +
        2\ii
        \tilde{G}_j
        \right)
        \right)
        \Bigg]\nonumber
        \\
        & \nonumber\\
        &=
        \frac{1}{2^{N-1}}
        \sum_{\substack{\mu_1,\mu_1'= \pm \\ \dots \\ \mu_N,\mu_N'= \pm}}
        \Bigg[
        e^{-2W(\Lambda_i+\Lambda_j,
        \Lambda_i+\Lambda_j)
        }
        \cos\left(
        2
        \tilde{G}_i
        +
        2
        \tilde{G}_j
        \right)\nonumber
        \\
        &
        ~~~~~~~~~~~~~~~~~~~~~~
        +
        e^{
        -2W(\Lambda_i-\Lambda_j,
        \Lambda_i-\Lambda_j)
        }
        \cos\left(
        2
        \tilde{G}_i
        -
        2
        \tilde{G}_j
        \right)
        \Bigg].\nonumber
\end{align}
It will be useful for later analysis to convert the sum over the eigenvalues to a product of cosines by pulling out the first term of the inner sum:
\begin{align}
        \sum_{\substack{\mu_1,\mu_1'= \pm \\ \dots \\ \mu_N,\mu_N'= \pm}}
        \Bigg[
        \cos\left(
        2\tilde{G}_i\pm2\tilde{G}_j
        \right)
        \Bigg]
        &=
        \sum_{\substack{\mu_1,\mu_1'= \pm \\ \dots \\ \mu_N,\mu_N'= \pm}}
        \Bigg[
        \cos\left(
        2\sum_{\substack{k=1\\ k \neq i,j}}^N
        \mu_k G_{ik}
        \pm
        2\sum_{\substack{k=1\\ k\neq i,j}}^N
        \mu_k G_{jk}
        \right)
        \Bigg] \\
        &=
        \sum_{\substack{\mu_1,\mu_1'= \pm \\ \dots \\ \mu_N,\mu_N'= \pm}}
        \Bigg[
        \cos\left(
        2\mu_1G_{i,1}
        \pm
        2\mu_1G_{j,1}
        +
        2\sum_{\substack{k=2\\k  \neq i,j}}^N
        \mu_k G_{ik}
        \pm
        2\sum_{\substack{k=2\\ \neq i,j}}^N
        \mu_k G_{jk}
        \right)
        \Bigg] \nonumber\\
        &=
        \sum_{\substack{\mu_1,\mu_1'= \pm \\ \dots \\ \mu_N,\mu_N'= \pm}}
        \Bigg[
        \cos\left(
        2\mu_1(G_{i,1}
        \pm
        G_{j,1})
        +
        2\sum_{\substack{k=2\\ k\neq i,j}}^N
        \mu_k G_{ik}
        \pm
        2\sum_{\substack{k=2\\ k\neq i,j}}^N
        \mu_k G_{jk}
        \right)
        \Bigg] \nonumber\\
        &=
        \sum_{\substack{\mu_2,\mu_2'= \pm \\ \dots \\ \mu_N,\mu_N'= \pm}}
        \Bigg[
        \cos\left(
        2(G_{i,1}
        \pm
        G_{j,1})
        +
        2\sum_{\substack{k=2\\ k \neq i,j}}^N
        \mu_k G_{ik}
        \pm
        2\sum_{\substack{k=2\\ k\neq i,j}}^N
        \mu_k G_{jk}
        \right)\nonumber\\
        &~~~~~~~~~~~~~~~+
        \cos\left(
        -2(G_{i,1}
        \pm
        G_{j,1})
        +
        2\sum_{\substack{k=2\\ k\neq i,j}}^N
        \mu_k G_{ik}
        \pm
        2\sum_{\substack{k=2\\ k\neq i,j}}^N
        \mu_k G_{jk}
        \right)
        \Bigg].\nonumber
\end{align}
We now use the two trigonometric identities 
\begin{equation}
    \cos(\alpha+\beta) = 
    \cos(\alpha)\cos(\beta)-\sin(\alpha)\sin(\beta)
\end{equation}
and
\begin{equation}
    \cos(\alpha-\beta) = 
    \cos(\alpha)\cos(\beta)+\sin(\alpha)\sin(\beta)
\end{equation}
to obtain
\begin{equation}
    \begin{aligned}
        \sum_{\substack{\mu_1,\mu_1'= \pm \\ \dots \\ \mu_N,\mu_N'= \pm}}
        \Bigg[
        \cos\left(
        2\tilde{G}_i\pm2\tilde{G}_j
        \right)
        \Bigg]
        &=
        2\cos\left(
        2G_{i,1}\pm 2G_{j,1}
        \right)
        \sum_{\substack{\mu_2,\mu_2'= \pm \\ \dots \\ \mu_N,\mu_N'= \pm}}
        \Bigg[
        \cos\left(
        2\sum_{\substack{k=2\\ k\neq i,j}}^N
        \mu_k G_{ik}
        \pm
        2\sum_{\substack{k=2\\ k\neq i,j}}^N
        \mu_k G_{jk}
        \right)
        \Bigg].
        \\
    \end{aligned}
\end{equation}
The technique above can also be used to factor out the remaining terms in the sum, yielding the product
\begin{equation}\label{pull out}
    \begin{aligned}
        \sum_{\substack{\mu_1,\mu_1'= \pm \\ \dots \\ \mu_N,\mu_N'= \pm}}
        \Bigg[
        \cos\left(
        2\tilde{G}_i \pm2\tilde{G}_j
        \right)
        \Bigg]
        &=
        2^{N-2}
        \prod_{\substack{k=1 \\ k\neq i,j}}^N
        \cos(2G_{ik}\pm 2G_{jk}).
    \end{aligned}
\end{equation}
The method outlined above will be repeatedly used to simplify sums in the following subsections.
The exponentials in Eq.~\eqref{evzz before pull out} can also be simplified by decomposing the Wightman function as
\begin{equation}
    \begin{aligned}
        W(f,g) &= 
        \frac{1}{2}
        \Big(
        H(f,g)+\ii E(f,g)
        \Big).
    \end{aligned}
\end{equation}
Using $W_{ij}\equiv \lambda^2 W(\Lambda_i,\Lambda_j)$ and use the decomposition we find
\begin{equation}
    \begin{aligned}
        -2\lambda^2 W(\Lambda_i+\Lambda_j,
        \Lambda_i+\Lambda_j) 
        &=
        -2(
        W_{ii}+W_{ij}
        +W_{ji}+W_{jj} )\\
        &=
        -(
        H_{ii}+\ii E_{ii} +
        H_{ij}+\ii E_{ij} +
        H_{ji}+\ii E_{ji} +
        H_{jj}+\ii E_{jj}) \\
        &=
        -(
        H_{ii}+H_{jj}+ 2H_{ij}
        ),
    \end{aligned}
\end{equation}
where the last line follows from the symmetry of $H$ and antisymmetry of $E$.
Similarly we have 
\begin{equation}
    \begin{aligned}
        -2\lambda^2 W(\Lambda_i-\Lambda_j,
        \Lambda_i-\Lambda_j) 
        &=
        -2(
        W_{ii}-W_{ij}
        -W_{ji}+W_{jj} )\\
        &=
        -(
        H_{ii}+\ii E_{ii} +
        -H_{ij}-\ii E_{ij} +
        -H_{ji}-\ii E_{ji} +
        H_{jj}+\ii E_{jj} )\\
        &=
        -(
        H_{ii}+H_{jj} -2H_{ij}
        ).
    \end{aligned}
\end{equation}
By inserting the result above and Eq.~\eqref{pull out} into Eq.~\eqref{evzz before pull out} we can recast the expected value as
\begin{align}\label{ev zz}
        \evv{\s_z^{(i)}\s_z^{(j)}}
        &=
        \frac{1}{2}\Bigg(
        e^{-H_{ii}-H_{jj}-2H_{ij}}
        \prod_{\substack{k=1 \\ k\neq i,j}}^N
        \cos(2G_{ik}+ 2G_{jk})
        +
        e^{-H_{ii}-H_{jj}+2H_{ij}}
        \prod_{\substack{k=1 \\ k\neq i,j}}^N
        \cos(2G_{ik}- 2G_{jk})
        \Bigg) \\
        &=
        \frac{1}{2}
        e^{-H_{ii}-H_{jj}}
        \Bigg(
        e^{+2H_{ij}}
        \prod_{\substack{k=1 \\ k\neq i,j}}^N
        \cos(2G_{ik}- 2G_{jk})
        +
        e^{-2H_{ij}}
        \prod_{\substack{k=1 \\ k\neq i,j}}^N
        \cos(2G_{ik}+ 2G_{jk})
        \Bigg).\nonumber
\end{align}

\subsection{Computing $\ev{\s_y^{(i)}\s_y^{(j)}}$}


Similar to the previous subsection, we start by studying the matrix elements \begin{equation}
    \s_y^{(i)}\s_y^{(j)}
    =
    \begin{pmatrix}
        0 & 0 & 0 & -1 \\
        0 & 0 & 1 & 0 \\
        0 & 1 & 0 & 0 \\
        -1 & 0 & 0 & 0 \\
    \end{pmatrix}.
\end{equation}
We immediately see the similarity to $\bra{\mu_i'\mu_j'}\s_z^{(i)}\s_z^{(j)}\ket{\mu_i\mu_j}$, with the only difference being the signs of the terms corresponding to $(\mu_i',\mu_j',\mu_i,\mu_j) = (\pm,\pm,\mp,\mp)$. This implies a similar expression to Eq.~\eqref{evzz first},  with the first and last term changing sign. Propagating this through the  calculation we find 
\begin{equation}\label{ev yy}
    \begin{aligned}
        \evv{\s_y^{(i)}\s_y^{(j)}}
        &=
        \frac{1}{2}
        e^{-H_{jj}-H_{jj}}
        \Bigg(
        e^{+2H_{ij}}
        \prod_{\substack{k=1 \\ k\neq i,j}}^N
        \cos(2G_{ik} - 2G_{jk})
        -
        e^{-2H_{ij}}
        \prod_{\substack{k=1 \\ k\neq i,j}}^N
        \cos(2G_{ik}+ 2G_{jk})
        \Bigg).
    \end{aligned}
\end{equation}

\subsection{Computing $\ev{\s_y^{(i)}\s_x^{(j)}}$ and $\ev{\s_y^{(j)}\s_x^{(i)}}$}
The matrix representation will in this case be
\begin{equation}\label{ev yx matrix}
    \s_y^{(i)}\s_x^{(j)}
    =
    \begin{pmatrix}
        0 & 0 & \ii & 0 \\
        0 & 0 & 0 & -\ii \\
        -\ii & 0 & 0 & 0 \\
        0 & \ii & 0 & 0 \\
    \end{pmatrix},
\end{equation}
giving us the table of eigenvalues
\begin{center}
    \begin{tabular}{| c c c c | c |} 
         \hline
         $\mu_i'$ & $\mu_j'$ 
         & $\mu_i$ & $\mu_j$
         & $\bra{\mu_i'\mu_j'}
         \s_y^{(i)}\s_x^{(j)}
         \ket{\mu_i\mu_j}$\\ [0.5ex] 
         \hline\hline
           + & + & - & + & $\,\,\ii$\\ 
         \hline
           + & - & - & - & $-\ii$\\
         \hline
         - & + & + & + & $-\ii$ \\
         \hline
           - & - & + & - & $\,\,\ii$\\ 
         \hline
    \end{tabular}.
\end{center}
By inserting these, term by term, into Eq.~\eqref{appeq:general ev useful} we obtain all non-zero terms in the sum.
We will simplify these using the methods and definitions developed in the previous subsections. Starting by pulling out the shared exponential we get
\begin{align}
        \evv{\s_y^{(i)}\s_x^{(j)}}_{\r_\tc{d}}
        &=
        \sum_{\substack{\mu_1,\mu_1'= \pm \\ \dots \\ \mu_N,\mu_N'= \pm}}
        \frac{\ii}{2^N}
        \Bigg[
        ~
        \exp\left(
        -\frac{\lambda^2}{2}
        ||
        2\Lambda_i
        ||^2
        \right)
        \exp
        \left(
        2
        \ii
        \tilde{G}_i
        +
        \ii \Delta_{ij}
        +
        \ii E_{ij}
        \right) \\
        &
        ~~~~~~~~~~~~~~~~~~
        -
        \exp\left(
        -\frac{\lambda^2}{2}
        ||
        -2\Lambda_i
        ||^2
        \right)
        \exp
        \left(
        2
        \ii
        \tilde{G}_i
        -
        \ii \Delta_{ij}
        -
        \ii E_{ij}
        \right) \nonumber\\
        &
        ~~~~~~~~~~~~~~~~~~
        -
        \exp\left(
        -\frac{\lambda^2}{2}
        ||
        2\Lambda_i
        ||^2
        \right)
        \exp
        \left(
        -2
        \ii
        \tilde{G}_i
        -
        \ii \Delta_{ij}
        -
        \ii E_{ij}
        \right) \nonumber\\
        &
        ~~~~~~~~~~~~~~~~~~
        +
        \exp\left(
        -\frac{\lambda^2}{2}
        ||
        -2\Lambda_i
        ||^2
        \right)
        \exp
        \left(
        -2
        \ii
        \tilde{G}_i
        +
        \ii \Delta_{ij}
        +
        \ii E_{ij}
        \right)
        \Bigg]\nonumber\\
        &=
        \sum_{\substack{\mu_1,\mu_1'= \pm \\ \dots \\ \mu_N,\mu_N'= \pm}}
        \frac{\ii}{2^N}
        e^{-2H_{ii}}
        \Bigg[
        \exp
        \left(
        2
        \ii
        \tilde{G}_i
        +
        \ii \Delta_{ij}
        +
        \ii E_{ij}
        \right) 
        -
        \exp
        \left(
        -2
        \ii
        \tilde{G}_i
        -
        \ii \Delta_{ij}
        -
        \ii E_{ij}
        \right) \nonumber
        \\
        &
        ~~~~~~~~~~~~~~~~~~~~~~~~~~~~~
        -
        \exp
        \left(
        2
        \ii
        \tilde{G}_i
        -
        \ii \Delta_{ij}
        -
        \ii E_{ij}
        \right)
        +
        \exp
        \left(
        -2
        \ii
        \tilde{G}_i
        +
        \ii \Delta_{ij}
        +
        \ii E_{ij}
        \right)
        \Bigg] \nonumber\\
        &=
        \sum_{\substack{\mu_1,\mu_1'= \pm \\ \dots \\ \mu_N,\mu_N'= \pm}}
        \frac{\ii}{2^N}
        e^{-2H_{ii}}
        \Bigg[
        \exp
        \left(
        2
        \ii
        \tilde{G}_i
        +
        2\ii G_{ij}
        \right) 
        -
        \exp
        \left(
        -2
        \ii
        \tilde{G}_i
        -
        2\ii G_{ij}
        \right) \nonumber
        \\
        &
        ~~~~~~~~~~~~~~~~~~~~~~~~~~~~~
        -
        \exp
        \left(
        2
        \ii
        \tilde{G}_i
        -
        2\ii G_{ij}
        \right)
        +
        \exp
        \left(
        -2
        \ii
        \tilde{G}_i
        +
        2\ii G_{ij}
        \right)
        \Bigg]\nonumber \\
        &=
        \sum_{\substack{\mu_1,\mu_1'= \pm \\ \dots \\ \mu_N,\mu_N'= \pm}}
        \frac{\ii}{2^{N-1}}
        e^{-2H_{ii}}
        \Bigg[
        \sinh
        \left(
        2
        \ii
        \tilde{G}_i
        +
        2\ii G_{ij}
        \right)  
        +
        \sinh
        \left(
        -2
        \ii
        \tilde{G}_i
        +
        2\ii G_{ij}
        \right)
        \Bigg]\nonumber \\
        &=
        \sum_{\substack{\mu_1,\mu_1'= \pm \\ \dots \\ \mu_N,\mu_N'= \pm}}
        \frac{-1}{2^{N-1}}
        e^{-2H_{ii}}
        \Bigg[
        \sin
        \left(
        2
        \tilde{G}_i
        +
        2 G_{ij}
        \right)  
        +
        \sin
        \left(
        -2
        \tilde{G}_i
        +
        2 G_{ij}
        \right)
        \Bigg].\nonumber
\end{align}
We will now use the trigonometric identities 
\begin{equation}
    \sin(\alpha+\beta)
    =
    \sin(\alpha)\cos(\beta)+\cos(\beta)\sin(\alpha)
\end{equation}
and
\begin{equation}
    \sin(\alpha-\beta)
    =
    \sin(\alpha)\cos(\beta)-\cos(\beta)\sin(\alpha)
\end{equation}
to obtain
\begin{equation}
    \begin{aligned}
        \evv{\s_y^{(i)}\s_x^{(j)}}_{\r_\tc{d}}
        &=
        \sum_{\substack{\mu_1,\mu_1'= \pm \\ \dots \\ \mu_N,\mu_N'= \pm}}
        \frac{-1}{2^{N-2}}
        e^{-2H_{ii}}
        \Bigg[
        \sin
        \left(
        2 G_{ij}
        \right)  
        \cos\left(
        2 \tilde{G}_{i}
        \right)
        \Bigg].
    \end{aligned}
\end{equation}
Using the pull out method discussed in Subsection~\ref{app:evzz}, we find
\begin{equation}\label{ev yx}
    \begin{aligned}
        \evv{\s_y^{(i)}\s_x^{(j)}}_{\r_\tc{d}}
        &=
        -
        e^{-2H_{ii}}
        \sin
        \left(
        2 G_{ij}
        \right)
        \prod_{\substack{k=1\\ k\neq ij}}^N
        \cos\left(
        2 G_{ik}
        \right).
    \end{aligned}
\end{equation}
The expected value $\langle{\s_x^{(i)}\s_y^{(j)}}\rangle = \langle{\s_y^{(j)}\s_x^{(i)}}\rangle$ can be obtained by permuting the labels in $\langle{\s_y^{(i)}\s_x^{(j)}}\rangle$. This gives us
\begin{equation}\label{ev xy}
    \begin{aligned}
        \evv{\s_x^{(i)}\s_y^{(j)}}_{\r_\tc{d}}
        &=
        -
        e^{-2H_{jj}}
        \sin
        \left(
        2 G_{ji}
        \right)
        \prod_{\substack{k=1\\ k\neq ij}}^N
        \cos\left(
        2 G_{jk}
        \right).
    \end{aligned}
\end{equation}

\subsection{Computing $\evv{\hat{\sigma}_z^{(i)}}$}

Let us start by noticing that $\langle \s_z^{(i)}\rangle = \langle{\s_z^{(i)}\mathds{I}^{(j)}}\rangle$. Writing the local expected value in this form, we can use the methods developed in the previous subsections. We obtain the matrix
\begin{equation}
    \s_z^{(i)}\mathds{I}^{(j)}
    =
    \begin{pmatrix}
        0 & 0 & 1 & 0 \\
        0 & 0 & 0 & 1 \\
        1 & 0 & 0 & 0 \\
        0 & 1 & 0 & 0 \\
    \end{pmatrix},
\end{equation}
giving us the non-zero matrix components 
\begin{center}
    \begin{tabular}{| c c c c | c|} 
         \hline
         $\mu_i'$ & $\mu_j'$ 
         & $\mu_i$ & $\mu_j$ 
         & $\bra{\mu_i'\mu_j'}
         \s_z^{(i)}\mathds{I}^{(j)}
         \ket{\mu_i\mu_j}$\\ [0.5ex] 
         \hline\hline
          + & + & - & + & 1\\ 
         \hline
          + & - & - & - & 1\\
         \hline
          - & + & + & + & 1\\
         \hline
          - & - & + & - & 1\\ 
         \hline
    \end{tabular}.
\end{center}
We immediately note that this structure is similar to that of Eq.~\eqref{ev yx matrix}. The only difference being the signs of two terms and the lack of the imaginary factor $\ii$. Since each term will be positive we will just get hyperbolic cosines where we previously got hyperbolic sines. From this we can see that 
\begin{equation}\label{ev z on alpha}
    \begin{aligned}
        \evv{\s_z^{(i)}\mathds{I}^{(j)}}
        &=
        e^{-2H_{ii}}
        \cos(2G_{ij})
        \prod_{\substack{k=1\\k\neq i,j}}^N
        \cos(2G_{ik}).
    \end{aligned}
\end{equation}
At the current stage one might think that the expression above has some special dependence on $j$, even though the operation is only on $i$ observable is local to system $i$. This is not the case , as the expression can be written as
\begin{equation}\label{z on alpha 2}
    \begin{aligned}
        \evv{\s_z^{(i)}}
        &=
        e^{-2H_{ii}}
        \prod_{\substack{k=1\\k\neq i}}^N
        \cos(2G_{ik}).
    \end{aligned}
\end{equation}

\section{}\label{Appendix List of nonzero expected values}
\subsection{Non-zero expected values}
The remaining expected values of up to two Pauli matrices not listed in the previous are all zero, apart from the identity on every detector, which is of course $1$. For convenience the nonzero expected values we have calculated are listed below:
\begin{equation}
    \begin{aligned}
        \evv{\s_z^{(i)}\s_z^{(j)}}
        &=
        \frac{1}{2}
        e^{-(H_{ii}+H_{jj})}
        \left(
        e^{2H_{ij}}\prod_{\substack{k=1 \\ k\neq i,j}}^N \cos(2G_{ik}-2G_{jk})
        +
        e^{-2H_{ij}}\prod_{\substack{k=1 \\ k\neq i,j}}^N \cos(2G_{ik}+2G_{jk})
        \right),
    \end{aligned}
\end{equation}

\begin{equation}
    \begin{aligned}
        \evv{\s_y^{(i)}\s_y^{(j)}}
        &=
        \frac{1}{2}
        e^{-(H_{ii}+H_{jj})}
        \left(
        e^{2H_{ij}}\prod_{\substack{k=1 \\ k\neq i,j}}^N \cos(2G_{ik}-2G_{jk})
        -
        e^{-2H_{ij}}\prod_{\substack{k=1 \\ k\neq i,j}}^N \cos(2G_{ik}+2G_{ik})
        \right),
    \end{aligned}
\end{equation}

\begin{equation}
    \begin{aligned}
        \evv{\s_z^{(i)}}
        &=
        e^{-H_{ii}}
        \cos(2G_{ij})
        \left(
        \prod_{\substack{k=1 \\ k\neq i,j}}^N \cos(2G_{ik})
        \right),
    \end{aligned}
\end{equation}

\begin{equation}
    \begin{aligned}
        \evv{\s_z^{(j)}}
        &=
        e^{-H_{jj}}
        \cos(2G_{ji})
        \left(
        \prod_{\substack{k=1 \\ k\neq i,j}}^N \cos(2G_{jk})
        \right),
    \end{aligned}
\end{equation}

\begin{equation}
    \begin{aligned}
        \evv{\s_y^{(i)}\s_x^{(j)}}
        &=
        e^{-H_{ii}}
        \sin(2G_{ji})
        \left(
        \prod_{\substack{k=1 \\ k\neq i,j}}^N \cos(2G_{ik})
        \right),
    \end{aligned}
\end{equation}

\begin{equation}
    \begin{aligned}
        \evv{\s_x^{(i)}\s_y^{(j)}}
        &=
        e^{-H_{jj}}
        \sin(2G_{ij})
        \left(
        \prod_{\substack{k=1 \\ k\neq i,j}}^N \cos(2G_{jk})
        \right).
    \end{aligned}
\end{equation}

\section{Extracting the Hadamard function for general separations}\label{Appendix Extracting the Hadamard function in genereal}
Our goal is now to extract the Hadamard function between pairs of detectors $i$ and $j$ from the non-zero expected values of Pauli observables. Once we write the smeared Hadamard function, an experimentalist could  extract the field's two-point function between all pairs of detectors from measurements on UDW-detectors.
To recover the smeared Hadamard function, we start by noticing that by adding Eq.~\eqref{ev zz} and Eq.~\eqref{ev yy} we obtain 
\begin{equation}
    \begin{aligned}
        \evv{\s_z^{(i)}\s_z^{(j)}}+
        \evv{\s_y^{(i)}\s_y^{(j)}}
        &=
        e^{-H_{ii}-H_{jj}}
        e^{+2H_{ij}}
        \prod_{\substack{k=1 \\ k\neq i,j}}^N
        \cos(2G_{ik}- 2G_{jk}),
    \end{aligned}
\end{equation}
and subtracting them yields
\begin{equation}
    \begin{aligned}
        \evv{\s_z^{(i)}\s_z^{(j)}}-
        \evv{\s_y^{(i)}\s_y^{(j)}}
        &=
        e^{-H_{ii}-H_{jj}}
        e^{-2H_{ij}}
        \prod_{\substack{k=1 \\ k\neq i,j}}^N
        \cos(2G_{ik}+ 2G_{jk}).
    \end{aligned}
\end{equation}
If we now take the ratio of these new expressions we have
\begin{equation}
    \begin{aligned}
        \frac{\evv{\s_z^{(i)}\s_z^{(j)}}+
        \evv{\s_y^{(i)}\s_y^{(j)}}
        }{\evv{\s_z^{(i)}\s_z^{(j)}}-
        \evv{\s_y^{(i)}\s_y^{(j)}}}
        &=
        e^{+4H_{ij}}
        \prod_{\substack{k=1 \\ k\neq i,j}}^N
        \frac{
        \left(
        1+\tan(2G_{ik})\tan(2G_{jk})
        \right)}
        {
        \left(
        1-\tan(2G_{ik})\tan(2G_{jk})
        \right)}.
    \end{aligned}
\end{equation}
Taking the natural logarithm on both sides and dividing by four gives us
\begin{align}
        \frac{1}{4}\ln\left(
        \frac{\evv{\s_z^{(i)}\s_z^{(j)}}+
        \evv{\s_y^{(i)}\s_y^{(j)}}
        }{\evv{\s_z^{(i)}\s_z^{(j)}}-
        \evv{\s_y^{(i)}\s_y^{(j)}}}
        \right)
        &=
        H_{ij}+
        \frac{1}{4}\ln\left(
        \prod_{\substack{k=1 \\ k\neq i,j}}^N
        \frac{
        \left(
        1+\tan(2G_{ik})\tan(2G_{jk})
        \right)}
        {
        \left(
        1-\tan(2G_{ik})\tan(2G_{jk})
        \right)}
        \right)
        \\
        &=
        H_{ij}+
        \frac{1}{2}
        \sum_{\substack{k=1 \\ k\neq i,j}}^N
        \frac{1}{2}
        \ln\left(
        \frac{
        \left(
        1+\tan(2G_{ik})\tan(2G_{jk})
        \right)}
        {
        \left(
        1-\tan(2G_{ik})\tan(2G_{jk})
        \right)}
        \right).\nonumber
\end{align}
If we now use the identity 
\begin{equation}
    \text{arctanh}(x) = 
    \frac{1}{2}\ln\left(\frac{1+x}{1-x}\right),
\end{equation}
we can simplify the expression to
\begin{equation}\label{Hab without evs}
    \begin{aligned}
        H_{ij}
        =
        \frac{1}{4}\ln\left( \frac{\evv{\s_z^{(i)}\s_z^{(j)}}+
        \evv{\s_y^{(i)}\s_y^{(j)}}
        }{\evv{\s_z^{(i)}\s_z^{(j)}}-
        \evv{\s_y^{(i)}\s_y^{(j)}}}\right)
        -
        \frac{1}{2}
        \sum_{\substack{k=1 \\ k\neq i,j}}^N
        \text{arctanh}\left(
        \tan(2G_{ik})\tan(2G_{jk})
        \right).
    \end{aligned}
\end{equation}
At this stage, while the first term is completely described by expected values, we still do not know how to evaluate the second term in terms of measurable quantities. We may however notice that the ratio of Eq.~\eqref{ev z on alpha} and Eq.~\eqref{ev yx} gives
\begin{equation}
    \begin{aligned}
        \frac{\evv{\s_y^{(i)}\s_x^{(j)}}}{\evv{\s_z^{(i)}}}
        &=
        \frac{ -
        e^{-2H_{ii}}
        \sin
        \left(
        2 G_{ij}
        \right)
        \prod_{\substack{k=1\\ k\neq i,j}}^N
        \cos\left(
        2 G_{ik}
        \right)
        }
        {
        e^{-H_{ii}}
        \cos(2G_{ij})
        \prod_{\substack{k=1\\ k\neq i,j}}^N
        \cos(2G_{ik})
        }\\
        &=
        -\tan(2G_{ij}).
    \end{aligned}
\end{equation}
Recall that we could also write Eq.~\eqref{ev z on alpha} as Eq.~\eqref{z on alpha 2}. We then see that 
\begin{equation}
    \begin{aligned}
        \frac{\evv{\s_y^{(i)}\s_x^{(k)}}}{\evv{\s_z^{(i)}}}
        &=
        -\tan(2G_{ik}).
    \end{aligned}
\end{equation}
Similarly,
\begin{equation}
    \begin{aligned}
        \frac{\evv{\s_x^{(k)}\s_y^{(j)}}}{\evv{\s_z^{(j)}}}
        &=
        -\tan(2G_{jk}).
    \end{aligned}
\end{equation}
We then have all the previously unknown terms in the right hand side of equation Eq.~\eqref{Hab without evs}. We may now write it in terms of expected values as
\begin{equation}
    \begin{aligned}
        H_{ij}
        =
        \frac{1}{4}\ln\left(\frac
        {\evv{\s_z^{(i)}\s_z^{(j)}}+
        \evv{\s_y^{(i)}\s_y^{(j)}}}
        {\evv{\s_z^{(i)}\s_z^{(j)}}-
        \evv{\s_y^{(i)}\s_y^{(j)}}}\right)
        -
        \frac{1}{2}
        \sum_{\substack{k=1 \\ k\neq i,j}}^N
        \text{arctanh}
        \left(
        \frac{\evv{\s_y^{(i)}\s_x^{(k)}}}{\evv{\s_z^{(i)}}}
        \frac{\evv{\s_x^{(k)}\s_y^{(j)}}}{\evv{\s_z^{(j)}}}
        \right).
    \end{aligned}
\end{equation}

\section{Details for error estimation}\label{Appendix Details for error estimation}
In this appendix we will provide the details for estimating the error in the estimation of the field's two-point function incurred from using physical detectors with finite sized interactions in spacetime as opposed to sharply localized interaction regions. Specifically, if the interaction regions are centered at events $\mf x_i$ and $\mf x_j$ with spacetime smearing functions $\Lambda_{\mf x_i}$ and $\Lambda_{\mf x_j}$ have characteristic length $\ell$ that controls their extension in both space and time, we are looking for an expression of the form
\begin{equation}
    W(\Lambda_{\mf x_i},\Lambda_{\mf x_j}) = 
    W\left(\Lambda_{\mf x_i}\eval_{\text{pointlike}}, \Lambda_{\mf x_j}\eval_{\text{pointlike}}\right)
    +
    \order{\ell^2}.
\end{equation}
We will obtain such an expression by performing a multipole expansion of the Wightman function. Such an expression can be obtained by considering a Taylor expansion of the Wightman function around events $\mf x_i$ and $\mf x_j'$ in Riemann normal coordinates (RNC) centered around these points. The expansion reads 
\begin{equation}
    W(\mf x,\mf x') = 
    \sum_{\substack{n=0 \\ m=0}}^\infty
    \frac{1}
    {n!m!}
    \pd{}{x^{\mu_1}}\dots\pd{}{x^{\mu_n}}
    \pd{}{x^{\nu_1'}}\dots\pd{}{x^{\nu_m'}}
    W(\mf x, \mf x')\eval_{\substack{\mf x = \mf x_i \\ \mf x'= \mf x_j}}
    (\mf x- \mf x_i)^{\mu_1}\dots
    (\mf x- \mf x_i)^{\mu_n}
    (\mf x'- \mf x_j')^{\nu_1'}\dots
    (\mf x'- \mf x_j')^{\nu_m'},
\end{equation}
where $(\mf x- \mf x_i)^{\mu}$ denote the RNCs of the event $\mf x$ centered at $\mf x_i$ and $(\mf x'- \mf x_j)^{\mu'}$ are the RNCs of $\mf x'$ centered at $\mf x_j$. Inserting the expansion above in the explicitl spacetime integrals for $W(\Lambda_{\mf x_i},\Lambda_\mf{x_j}$) yields
\begin{equation}\label{}
    \begin{aligned}
        W(\Lambda_1,\Lambda_2) 
        &= 
        \sum_{\substack{n=0 \\ m=0}}^\infty
        \frac{1}
        {n!m!}
        \pd{}{x^{\mu_1}}\dots\pd{}{x^{\mu_n}}
        \pd{}{x^{\nu_1'}}\dots\pd{}{x^{\nu_m'}}
        W(\mf x, \mf x')\eval_{\substack{\mf x = \mf x_0 \\ \mf x'= \mf x_0'}}
        \\
        &
        \int \dd^4 \mf x
        (\mf x- \mf x_0)^{\mu_1}\dots
        (\mf x- \mf x_0)^{\mu_n}
        \Lambda_1(\mf x)
        \int \dd^4 \mf x'
        (\mf x'- \mf x_0')^{\nu_1}\dots
        (\mf x'- \mf x_0')^{\nu_m}
        \Lambda_2(\mf x').
    \end{aligned}
\end{equation}
To simplify expressions, we define 
\begin{equation}
    \begin{aligned}
        W_{\mu_1\dots\mu_n\nu_1'\dots\nu_m'}
        (\mf x_j, \mf x_j)
        \equiv
        \pd{}{x^{\mu_1}}\dots\pd{}{x^{\mu_n}}
        \pd{}{x^{\nu_1'}}\dots\pd{}{x^{\nu_m'}}
        W(\mf x, \mf x')\eval_{\substack{\mf x = \mf x_i \\ \mf x'= \mf x_j}}
    \end{aligned}
\end{equation}
and the the spacetime multipole moments of the smearings functions around the points $\mf x_i$ and $\mf x_j$
\begin{equation}\label{Lambda moment appendix}
    \begin{aligned}
        \Lambda^{\mu_1\dots\mu_n}_{\mf x_i}
        \equiv
        \int \dd V (\mf x-\mf x_i)^{\mu_1}
        \dots(\mf x-\mf x_i)^{\mu_n}
        \Lambda_{\mf x_i}(\mf x)
    \end{aligned}
\end{equation}
and
\begin{equation}
    \begin{aligned}
        \Lambda^{\nu_1'\dots\nu_m'}_{\mf x_j}
        \equiv
        \int \dd V' (\mf x'-\mf x_j)^{\nu_1'}
        \dots(\mf x'-\mf x_j)^{\nu_m'}
        \Lambda_{\mf x_j}(\mf x').
    \end{aligned}
\end{equation}
With this notation we can write the smeared Wightman function as 
\begin{equation}\label{Wightman expanded appendix}
    \begin{aligned}
        W(\Lambda_{\mf x_i},\Lambda_{\mf x_j}) =
        \sum_{\substack{n=0 \\ m=0}}
        \frac{1}{n!m!}
        W_{\mu_1\dots\mu_n\nu_1'\dots\nu_m'}
        (\mf x_i,\mf x_j)
        \Lambda^{\mu_1\dots\mu_n}_{\mf x_i}
        \Lambda^{\nu_1'\dots\nu_m'}_{\mf x_j}.
    \end{aligned}
\end{equation}

We will choose the spacetime smearing functions to be of the form
We will choose the spacetime smearing functions $\Lambda_{\mf x_i}(\mf x)$ to be Gaussians when written in RNC centered at each respective point $\mf x_i$, such that the detectors interaction regions will be 4 dimensional hyperspheres in spacetime. 
Explicitly, denoting $\mf y^\mu=(\mf x-\mf x_0)^\mu$, the spacetime smearing functions will be prescribed as
\begin{equation}
    \Lambda_{\mf x_i}(\mf x)
    =
    \frac{
    e^{\frac{-(\mf y^0)^2-\mf y^i\mf y^j\delta_{ij}}{2\ell^2}}}
    {(2\pi)^2\ell^4}.
\end{equation}
Since the detector is Gaussian it is also even, thus the odd moments vanish for each detector. What this means is that we will have to change both the volume element and the separation factors in the integrals. Given that we are interested in the small $\ell$ limit, we can also expand the volume element to second order in RNC~\cite{poisson}:
\begin{equation}\label{dV}
    \dd V \rightarrow \dd^4\mf y\sqrt{-g}
    =
    \dd^4\mf y \Big(1-\frac{1}{6}R_{\alpha\beta}(\mf x_0)\mf y^\alpha \mf y^\beta
    \Big).
\end{equation}
In the next subsections we will evaluate the relevant moments of the spacetime smearing functions $\Lambda_{\mf x_i}$.

\subsection{Multipole moments in general spacetimes}

Let us start by defining the multipoles of the spacetime smearing functions integrated in Riemann normal coordinates without, incorporating the leading order expansion of the volume element of Eq.~\eqref{dV}. We call these quantities the \textit{flat} multipoles of $\Lambda_{\mf x_i}$. For instance, the flat monopole, dipole, and quadrupole are given by
\begin{equation}\label{Def bar I}
    \begin{aligned}
        I_{\mf x_i}
        \equiv 
        \int \dd^4 \mf y ~ \Lambda_{\mf x_i}(\mf y), \quad\quad I^\mu_{\mf x_i}
        \equiv 
        \int \dd^4 \mf y ~ \mf y^\mu \Lambda_{\mf x_i}(\mf y), \quad\quad I_{\mf x_i}^{\mu\nu}
        \equiv 
        \int \dd^4 \mf y ~ \mf y^\mu \mf y^\nu
        \Lambda_{\mf x_i}(\mf y).
    \end{aligned}
\end{equation}
Due to $\Lambda_{\mf x_i}(\mf y)$ being a normalized Gaussian, we see that $I_{\mf x_i}=1$. Due to symmetry, $I_{\mf x_i}^\mu = 0$, as well as the off-diagonal elements of $I_{\mf x_i}^{\mu\nu}$. The $00$ component is given by
\begin{align}\label{monopole calc}
        I_{x_i}^{00}
        &=\frac{1}{(2\pi\ell^2)^2}
        \int\dd^4 \mf y~
        (\mf y^{0})^2
        e^{-\frac{(\mf y^0)^2}{2\ell^2}}
        e^{-\frac{\mf y^i \mf y^j\delta_{ij}}{2\ell^2}}\\
        &=
        \frac{1}{(2\pi\ell^2)^2}
        \int\dd \mf y^0~
        (\mf y^{0})^2
        e^{-\frac{(\mf y^0)^2}{2\ell^2}}
        \int \dd^3\mf y
        e^{-\frac{\mf y^i \mf y_i}{2\ell^2}}
        \\
        &=
        \frac{1}{(2\pi\ell^2)^2}
        \int\dd \mf y^0~
        (\mf y^{0})^2
        e^{-\frac{(\mf y^0)^2}{2\ell^2}}
        \left(
        \int \dd y
        e^{-\frac{y^2}{2\ell^2}}
        \right)^3\nonumber
        \\
        &=
        \frac{1}{(2\pi\ell^2)^2}
        \left(
        \frac{\sqrt{\pi}(\sqrt{2}\ell)^3}
        {2}
        \right)
        \left(
        \sqrt{2\pi}\ell
        \right)^3\nonumber
        \\
        &=
        \frac{2^3\pi^2\ell^6}
        {2^3\pi^2\ell^4}\nonumber
        \\
        &=
        \ell^2.\nonumber
\end{align}
Notice that other diagonal elements will also yield the same result, so that
\begin{equation}
    I_{\mf x_i}^{\mu\nu}=\ell^2\delta^{\mu\nu}.
\end{equation}

The monopole moment~\eqref{Lambda moment appendix}, to leading order in $\ell$ is then found by incorporating the volume element expansion of Eq.~\eqref{dV}:
\begin{align}
        \Lambda_{\mf x_0}
        &=
        \int\dd V \Lambda(\mf x)
        \\
        &=
        I_{\mf x_0} -
        \frac{1}{6}R_{\alpha\beta}(\mf x_0)
        I^{\alpha\beta}_{\mf x_0}
        +
        \order{\ell^4}\nonumber
        \\
        &=
        1-\frac{1}{6}\ell^2
        R_{\alpha\beta}(\mf x_0)
        \delta^{\alpha\beta}
        +
        \order{\ell^4}\nonumber.
\end{align}
Also notice that the dipole moment will vanish, as it will always contain an odd power of the components of $\mf y^\mu$ in RNC. The quadrupole moment $\Lambda_{\mf x_i}^{\mu\nu}$ is also found by using Eq.~\eqref{Lambda moment appendix} and Eq.~\eqref{dV}, yielding
\begin{align}
        \Lambda_{\mf x_i}^{\mu\nu}
        &=
        \int \dd V ~
        \mf y^\mu
        \mf y^\nu
        \Lambda_{\mf x_i}(\mf y)
        \\
        &=
        \int
        \dd^4 \mf y~
        (1-\frac{1}{6}R_{\alpha\beta}(\mf x_0)
        \mf y^\alpha \mf y^\beta)
        \mf y^\mu \mf y^\nu
        \Lambda(\mf y)\nonumber
        \\
        &=
        I^{\mu\nu}_{\mf x_i}
        + \order{\ell^4}\nonumber
        \\
        &=
        \ell^2\delta^{\mu\nu}
        + \order{\ell^4}\nonumber.
\end{align}

\subsection{Leading order multipole expansion of the Wightman function}

Let us now use Eq.~\eqref{Wightman expanded appendix} to compute the leading order terms of multipole expansion of the smeared Wightman function. The leading order multipole expansion is
\begin{align}
        W(\Lambda_{\mf x_i},\Lambda_{\mf x_j})
        &=
        W(\mf x_i, \mf x_j)
        \Lambda_{\mf x_i}
        \Lambda_{\mf x_j}
        \\
        &
        +
        W_{\alpha}(\mf x_i, \mf x_j)
        \Lambda^{\alpha}_{\mf x_i}
        \Lambda_{\mf x_j}
        +
        W_{\mu'}(\mf x_i, \mf x_j)
        \Lambda_{\mf x_i}
        \Lambda^{\mu'}_{\mf x_j}\nonumber
        \\
        &+
        \frac{1}{2}
        W_{\alpha\beta}(\mf x_i, \mf x_j)
        \Lambda^{\alpha\beta}_{\mf x_i}
        \Lambda_{\mf x_j}
        +
        \frac{1}{2}
        W_{\mu'\nu'}(\mf x_i, \mf x_j)
        \Lambda_{\mf x_i}
        \Lambda^{\mu'\nu'}_{\mf x_j}\nonumber
        \\
        &+
        W_{\alpha\mu'}(\mf x_i, \mf x_j)
        \Lambda^{\alpha}_{\mf x_i}
        \Lambda^{\mu'}_{\mf x_j}.\nonumber
\end{align}

Inserting the values for the poles we found in the previous subsections give us

\begin{align}\label{Appendixeq error estimate}
        W(\Lambda_{\mf x_i},\Lambda_{\mf x_j})
        &=
        W(\mf x_i, \mf x_j)
        \left(
        1-\frac{1}{6}\ell^2
        R_{\alpha\beta}(\mf x_i)
        \delta^{\alpha\beta}
        +
        \order{\ell^4}
        \right)
        \left(
        1-\frac{1}{6}\ell^2
        R_{\alpha'\beta'}(\mf x_j)
        \delta^{\alpha'\beta'}
        +
        \order{\ell^4}
        \right)
        \\
        &
        +
        0
        +
        0\nonumber
        \\
        &+
        \frac{1}{2}
        W_{\mu\nu}(\mf x_i, \mf x_j)
        \left(
        \ell^2\delta^{\mu\nu}
        + \order{\ell^4}
        \right)
        \left(
        1-\frac{1}{6}\ell^2
        R_{\alpha'\beta'}(\mf x_j)
        \delta^{\alpha'\beta'}
        +
        \order{\ell^4}
        \right)\nonumber
        \\
        &
        +
        \frac{1}{2}
        W_{\mu'\nu'}(\mf x_i, \mf x_j')
        \left(
        1-\frac{1}{6}\ell^2
        R_{\alpha\beta}(\mf x_i)
        \delta^{\alpha\beta}
        +
        \order{\ell^4}
        \right)
        \left(
        \ell^2\delta^{\mu'\nu'}
        + \order{\ell^4}
        \right)\nonumber
        \\
        &+
        0\nonumber\\
        &=
        W(\mf x_i,\mf x_j)\nonumber
        \\
        &+
        \frac{\ell^2}{2}
        \left(
        -
        \frac{1}{6}
        W(\mf x_i,\mf x_j)
        R_{\alpha\beta}(\mf x_i)\delta^{\alpha\beta}
        -
        \frac{1}{6}
        W(\mf x_i,\mf x_j)
        R_{\alpha'\beta'}(\mf x_j)\delta^{\alpha'\beta'}
        +
        W_{\mu\nu}(\mf x_i, \mf x_j)
        \delta^{\mu\nu}
        +
        W_{\mu'\nu'}(\mf x_i, \mf x_j)
        \delta^{\mu'\nu'}\nonumber
        \right)
        \\
        &+
        \order{\ell^4}.\nonumber
\end{align}
We see that the leading order effect of using physical detectors is of order $\ell^2$.

\subsection{Minkowski Spacetime}\label{app:Mink}
To estimate the error in Minkowski spacetime we will use the fact that the kernel to the Wightman distribution may be written as
\begin{equation}\label{Appendixeq Wightman Minkowski}
    W(\mf x, \mf x')=
    \frac{1}{4\pi^2\big(-(t-t')^2+(\bar{x}-\bar{x}')^2\big)},
\end{equation}
We also note that the first two terms of order $\ell^2$ in Eq.~\eqref{Appendixeq error estimate} vanish since the Ricci tensor vanish in Minkowski spacetime. To obtain the final two terms we need to differentiate Eq.~\eqref{Appendixeq Wightman Minkowski} twice. Defining $\sigma = \tfrac{1}{2}(\mf x -\mf x')^\mu (\mf x -\mf x')_\mu$, the first derivative reads
\begin{align}
        W_{\mu}(\mf x, \mf x') 
        &=
        \pd{}{x^{\mu}}
        \frac{1}{8\pi^2\sigma} =
        -\frac{1}{8\pi^2\sigma^2} \partial_\mu \sigma = -\frac{(\mf x- \mf x')_{\mu}}{8\pi^2\sigma^2},\nonumber
\end{align}
where we used $\partial_\mu \sigma = (\mf x - \mf x')_\mu$.

The second derivatives are obtained by differentiating once more,
\begin{align}
        W_{\mu\nu}(\mf x, \mf x')
        &=
        -\pd{}{\mf x^{\nu}}
        \frac{(\mf x- \mf x')_{\mu}}{8\pi^2\sigma^2}
        \\
        &=
        -(\mf x- \mf x')_{\mu}\pd{}{\mf x^{\nu}}
        \frac{1}{8\pi^2\sigma^2}
        -
        \frac{1}{8\pi^2\sigma^2}
        \pd{}{\mf x^{\nu}}
        (\mf x- \mf x')_{\mu}
        \nonumber
        \\
        &=
        (\mf x- \mf x')_{\mu}
        \frac{1}{4\pi^2\sigma^3}
       \partial_\nu\sigma
        -
        \frac{1}{8\pi^2\sigma^2}
        \eta_{\mu\nu}
        \nonumber
        \\
        &=
        \frac{
        (\mf x- \mf x')_{\mu}(\mf x- \mf x')_{\nu}}
        {4\pi^2\sigma^3}
        -
        \frac{1}{8\pi^2\sigma^2}
        \eta_{\mu\nu}
        \nonumber
        \\
        &=
        \frac{W(\mf x, \mf x')}{\sigma}
        \left(
        \frac{2(\mf x- \mf x')_{\mu}(\mf x- \mf x')_{\nu}}{\sigma}
        - \eta_{\mu\nu}
        \right)
        \nonumber
\end{align}
Due to the symmetry $\mf x \leftrightarrow \mf x'$, we also have
\begin{equation}
    W_{\mu'\nu'}(\mf x, \mf x') =
    \frac{W(\mf x, \mf x')}{\sigma}
    \left(
    \frac{2(\mf x- \mf x')_{\mu'}(\mf x- \mf x')_{\nu'}}{\sigma}
    - \eta_{\mu'\nu'}
    \right)
\end{equation}

Using the results above in Eq.~\eqref{Appendixeq error estimate}, we find the expansion
\begin{align}
        W(\Lambda_{\mf x_i},\Lambda_{\mf x_j})
        &=
        W(\mf x_i, \mf x_j)
        \\
        &+
        \frac{\ell^2}{2}\!
        \left(
        \frac{W(\mf x_i, \mf x_j)}{\sigma}
        \left(
        \frac{2(\mf x_i-\mf x_j)_{\mu}(\mf x_i-\mf x_j)_{\nu}}{\sigma}
        -
        \eta_{\mu\nu}\!
        \right)
        \delta^{\mu\nu}
        +
        \frac{W(\mf x_i, \mf x_j)}{\sigma}
        \!
        \left(
        \frac{2(\mf x_i-\mf x_j)_{\mu'}(\mf x_i-\mf x_j)_{\nu'}}{\sigma}
        -
        \eta_{\mu'\nu'}\!
        \right)\!
        \delta^{\mu'\nu'}\!
        \right)\nonumber
        \\
        &+
        \order{\ell^4}\nonumber
        \\
        &=
        W(\mf x_i, \mf x_j)\nonumber
        \\
        &+
        \ell^2 W(\mf x_i, \mf x_j)
        \left(\!\!
        \left(
        \frac{4(\mf x_i-\mf x_j)_{\mu}(\mf x_i-\mf x_j)_{\nu}\delta^{\mu\nu}-4 \sigma}{4\sigma^2}
        \right)
        +
        \left(
        \frac{4(\mf x_i-\mf x_j)_{\mu'}(\mf x_i-\mf x_j)_{\nu'}\delta^{\mu'\nu'}-4\sigma}{4\sigma^2}
        \right)\!\!
        \right)+\order{\ell^4}\nonumber
        \\
        &=
        W(\mf x_i, \mf x_j)
        \left(
        1
        +
        \ell^2
        \left(
        \frac{4(\Delta t_{ij}^2+\Delta x_{ij}^2)- 2(-\Delta t_{ij}^2+\Delta x_{ij}^2)}{(-\Delta t^2+\Delta x_{ij}^2)^2}
        \right)
        \right)+\order{\ell^4}\nonumber
        \\
        &=
        W(\mf x_i, \mf x_j)
        \left(
        1
        +
        \ell^2
        \left(
        \frac{6\Delta t_{ij}^2+2\Delta x_{ij}^2}{(-\Delta t^2+\Delta x_{ij}^2)^2}
        \right)
        \right)+\order{\ell^4}\nonumber,
\end{align}
where $\Delta t_{ij} = |t_i - t_j|$ and $\Delta x_{ij} = |\bm x_i - \bm x_j|$.
When the regions are separated only spatially ($\Delta t_{ij}=0$) we then get
\begin{align}
    W(\Lambda_{\mf x_i},\Lambda_{\mf x_j})&=
    W(\mf x_i, \mf x_j)
    \left(
    1
    +
    \frac{4\ell^2}{\Delta x_{ij}^2}
    \right)+\order{\ell^4} =
    \left(
    \frac{1}{4\pi^2\Delta x_{ij}^2}
    +
    \frac{\ell^2}{\pi^2\Delta x_{ij}^4}
    \right)+\order{\ell^4} 
\end{align}
and for purely temporal separations ($\Delta t_{ij}=0$) we get
\begin{align}
    W(\Lambda_{\mf x_i},\Lambda_{\mf x_j})&=
    W(\mf x_i, \mf x_j)
    \left(
    1
    +
    \frac{12\ell^2}{\Delta t_{ij}^2}
    \right)+\order{\ell^4}
    =
    -\left(
    \frac{1}{4\pi^2\Delta t_{ij}^2}
    +
    \frac{3\ell^2}{\pi^2\Delta t_{ij}^4}
    \right)
    +\order{\ell^4}.
\end{align}

\section{The two-point function of a one-particle Gaussian wavepacket}\label{app: One particle state}

In this brief appendix we will compute the two-point function of a one-particle wavepacket of a massless scalar field with the specific Gaussian momentum distribution in Eq.~\eqref{eq:f(k)}  in 3+1 Minkowski spacetime. Explicitly, we will consider the wave-packet
\begin{equation}\label{apeq:coh gaussian}
    \ket{\varphi} = \int \dd^3\bm k f(\bm{k}) \hat{a}_{\bm k}^\dagger \ket{0}, \quad \text{with} \quad f(\bm k) =
    \frac{\delta^2}{\sqrt{4\pi}}
    \sqrt{2 |\bm{k}|} 
    e^{-\frac{\delta^2 \bm k^2}{2}},
\end{equation}
with $\delta$ being a constant determining the spread over wave-vectors. The two-point function is given by Eq.~\eqref{eq:tilde W}, so it is enough to find the functions $F(\mf x)$ explicitly:
\begin{align}
        F(\mf x)&= \int \dd^3 \bm k ~
        u_{\bm k}(\mf x) f(\bm k)\\\
        &=
        \int \dd^3 \bm k~
        \frac{1}{(2\pi)^{3/2}\sqrt{2 k}} e^{-\ii |\bm k| t + \bm k\cdot \bm x}
        \frac{\delta^2}{\sqrt{4\pi}}
        N \sqrt{2 |\bm{k}|} 
        e^{-\frac{\delta^2 |\bm k|^2}{2}}\nonumber
        \\
        &=
        \frac{
        e^{-\frac{(|\bm x|+t)^2}{2 \delta ^2}} 
        \left(
        e^{\frac{2 |\bm x| t}{\delta ^2}}
        (|\bm x|-t)  
        \left(1+\ii \text{erfi}\left(\frac{|\bm x|-t}{\sqrt{2} \delta }\right)
        \right)
        +
        (|\bm x|+t) 
        \left(1-\ii \text{erfi}\left(\frac{|\bm x|+t}{\sqrt{2} \delta }\right)\right)\right)
        }
        {
        4 \sqrt{\pi } \delta  |\bm x|
        },\nonumber
\end{align}
with complex conjugate
\begin{equation}
    F^*(\mf x)
    =
    \frac{e^{-\frac{(|\bm x|+t)^2}{2 \delta ^2}} \left((|\bm x|-t) e^{\frac{2 |\bm x| t}{\delta ^2}} \left(1-\ii \text{erfi}\left(\frac{|\bm x|-t}{\sqrt{2} \delta }\right)\right)+(|\bm x|+t) \left(1+\ii \text{erfi}\left(\frac{|\bm x|+t}{\sqrt{2} \delta }\right)\right)\right)}{4 \sqrt{\pi } \delta  |\bm x|}.
\end{equation}
We may now use Eq.~\eqref{eq:tilde W} and Eq.~\eqref{error estimate} to obtain the multipole expansion of the Wightman function. Since we are once more considering a case where the Ricci scalars vanish we only need to consider two terms of the corrections. The resulting expression is fairly unhinged and not particularly informative, so we display the resulting multipole expansion in Fig.~\ref{fig:1 particle} in the main text.

\section{The two-point function of a Gaussian sourced coherent state}\label{app:coherent}

In this appendix we will compute the two-point function of a coherent state sourced by a Gaussian test function in a massless scalar field in 3+1 dimensional Minkowski spacetime. A coherent state is defined by a classical solution $\phi_0(\mf x)$, which can be parametrized by a test function $g_0\in C^\infty_0(\M)$ such that $\phi_0 = Eg_0$. One can write the coherent state associated to the classical solution $\phi_0$ as $\ket{g_0} = e^{-\ii \hat{\phi}(g_0)}\ket{0}$. Noticing that  $e^{\ii \hphi(f_0)}\hphi(f)e^{-\ii \hphi(f_0)} = \hphi(f) + \phi_0(f)\mathds{I}$,
we can compute the two-point function of the coherent state as 
\begin{align}\label{appeq:smeared coh W}
    \bra{g_0}\hphi(\Lambda_i)\hphi(\Lambda_j)\ket{g_0} &= \bra{0}(\hphi(\Lambda_i)+\phi(\Lambda_i)) (\hphi(\Lambda_j)+\phi(\Lambda_j)\ket{0}
    \\
    &=
    \langle\hphi(\Lambda_i)\hphi(\Lambda_j)\rangle_0+
    \phi_0(\Lambda_i)\langle\hphi(\Lambda_j)\rangle+\phi_0(\Lambda_j)\langle\hphi(\Lambda_i)\rangle_0+
    \phi_0(\Lambda_i)\phi_0(\Lambda_j).\nonumber
    \\
    &=
    \langle\hphi(\Lambda_i)\hphi(\Lambda_j)\rangle_0+\phi_0(\Lambda_i)\phi_0(\Lambda_j)
    \nonumber,
\end{align}
where in the last equality we used the fact that the vacuum odd point functions vanish.

To analyze an explicit example, we choose the Gaussian test function
\begin{equation}
    g_0(\mf x) = \frac{e^{-\frac{t^2 + |\bm x|^2}{\delta^2}}}
    {(2\pi^2)^{3/2}\delta^3},
\end{equation}
where $\delta$ is a parameter with units of length that controls the effective size of the support of $g_0(\mf x)$. To compute the associated classical solution $\phi_0(\mf x)$, we use the fact that the causal propagator smeared against two Gaussians can be computed in closed form (see e.g.~\cite{perche2024closed}), and pick the auxiliary function 
\begin{equation}
    f_{\epsilon,\mf x_0}(\mf x) = \frac{e^{-\frac{(t-t_0)^2 + |\bm x - \bm x_0|^2}{2 \epsilon^2}}}
    {(2\pi^2)^2\epsilon^4},
\end{equation}
with the property that $\lim_{\epsilon\to 0} f_{\epsilon,\mf x_0}(\mf x) = \delta^{(4)}(\mf x - \mf x_0)$. We can then write the value of $\phi_0$ at each event $\mf x_0$ as
\begin{equation}
    \phi_0(\mf x_0) = \lim_{\epsilon \to 0} E(f_{\epsilon,\mf x_0},g_0).
\end{equation}
Adapting the results of~\cite{perche2024closed}, we find
\begin{align} 
    E(f_{\epsilon,\mf x_0},g_0) &=
    \frac{\delta  \left(e^{-\frac{(|\bm x_0|+t_0)^2}{4 \left(\delta^2+\epsilon ^2\right)}}-e^{-\frac{(|\bm x_0|-t_0)^2}{4 \left(\delta ^2+\epsilon ^2\right)}}\right)}{4 \sqrt{2} \pi  |\bm x_0| \sqrt{\delta ^2+\epsilon ^2}}.
\end{align}
Taking the limit of $\epsilon\to 0$ we then find
\begin{equation}\label{appeq:phi0x}
    \phi_0(\mf x)
    =
    \frac{e^{-\frac{(|\bm x|+t)^2}{4 \delta ^2}}-e^{-\frac{(|\bm x|-t)^2}{4 \delta ^2}}}{4 \sqrt{2} \pi  |\bm x|}.
\end{equation}
More generally, we can also compute $\phi_0(\Lambda_i)$ for the Gaussian spacetime functions $\Lambda_i$ used in Section~\ref{sec:examples}:
\begin{align}\label{appeq:phi0Lambda}
    \phi_0(\Lambda_i) = 
    E(\Lambda_i,g_0) &= 
    \frac{\delta  \left(e^{-\frac{(|\bm x_i|+t_i)^2}{4 \left(\delta ^2+\ell ^2\right)}}-e^{-\frac{(|\bm x_i|-t_i)^2}{4 \left(\delta ^2+\ell ^2\right)}}\right)}{4 \sqrt{2} \pi  |\bm x_i| \sqrt{\delta ^2+\ell ^2}}.
\end{align}
With Eqs.~\eqref{appeq:phi0x} and~\eqref{appeq:phi0Lambda}, we then obtain the two-point function and relevant smeared Wightman function of the Gaussian sourced coherent state.

\end{document}